\documentclass[twocolumn]{aastex701}

\usepackage{amsmath} 
\usepackage{color}
\usepackage{tabularx}

\begin{document}

\title{SN 2024dy: Dust formation in a long-lived Type IIn supernova and constraints on the dust mass}

\author{Sota Goto}
\affiliation{Graduate School of Science and Engineering, Kagoshima University, 1-21-35 Korimoto,Kagoshima, Kagoshima 890-0065, Japan}
\email[show]{k3110758@kadai.jp}
\email[show]{sota363636@gmail.com}

\author{Masayuki Yamanaka}
\affiliation{Amanogawa Galaxy Astronomy Research Center (AGARC), Graduate School of Science and Engineering, Kagoshima University, 1-21-35 Korimoto, Kagoshima, Kagoshima 890-0065, Japan}
\email{yamanaka@sci.kagoshima-u.ac.jp}

\author{Koji S. Kawabata}
\affiliation{Hiroshima Astrophysical Science Centre, Hiroshima University, 1-3-1 Kagamiyama, Higashi-Hiroshima, Hiroshima 739-8526, Japan}\affiliation{Department of Physics, Graduate School of Advanced Science and Engineering, Hiroshima University, 1-3-1 Kagamiyama, Higashi-Hiroshima, Hiroshima 739-8526, Japan}
\email{kawabtkj@hiroshima-u.ac.jp}

\author{Masaomi Tanaka}
\affiliation{Astronomical Institute, Tohoku University, Sendai, Miyagi 980-8578, Japan}
\email{masaomi.tanaka@astr.tohoku.ac.jp}

\author{Avinash Singh}
\affiliation{The Oskar Klein Centre, Department of Astronomy, Stockholm University, AlbaNova, SE-106 91 Stockholm, Sweden}
\email{avinash21292@gmail.com}  

\author{Devendra K. Sahu}
\affiliation{Indian Institute of Astrophysics, II Block, Koramangala, Bengaluru-560034, Karnataka, India}
\email{dks@iiap.res.in}

\author{Tatsuya Nakaoka}
\affiliation{Hiroshima Astrophysical Science Centre, Hiroshima University, 1-3-1 Kagamiyama, Higashi-Hiroshima, Hiroshima 739-8526, Japan}
\email{nakaokat@hiroshima-u.ac.jp}

\author{Anjasha Gangopadhyay}
\affiliation{The Oskar Klein Centre, Department of Astronomy, Stockholm University, AlbaNova, SE-106 91 Stockholm, Sweden}
\email{anjashagangopadhyay@gmail.com}

\author{Takahiro Nagayama}
\affiliation{Graduate School of Science and Engineering, Kagoshima University, 1-21-35 Korimoto,Kagoshima, Kagoshima 890-0065, Japan}
\email{nagayama@sci.kagoshima-u.ac.jp}

\author{Naveen Dukiya}
\affiliation{Aryabhatta Research Institute of Observational Sciences (ARIES), Manora Peak, Nainital-263002, India}\affiliation{Mahatma Jyotiba Phule Rohilkhand University, Pilibhit Bypass Road, Bareilly 243006, Uttar Pradesh, India}
\email{ndukiya@aries.res.in}
\author{Monalisa Dubey}
\affiliation{Aryabhatta Research Institute of Observational Sciences (ARIES), Manora Peak, Nainital-263002, India}\affiliation{Mahatma Jyotiba Phule Rohilkhand University, Pilibhit Bypass Road, Bareilly 243006, Uttar Pradesh, India}
\email{monalisa@aries.res.in}
\author{Kuntal Misra}
\affiliation{Aryabhatta Research Institute of Observational Sciences (ARIES), Manora Peak, Nainital-263002, India}
\email{[kuntal@aries.res.in}
\author{Bhavya Ailawadhi}
\affiliation{Physical Research Laboratory, Navrangpura, Ahmedabad, Gujarat-380009, India}
\email{}

\begin{abstract}
Type~IIn supernovae (SNe) are a subclass of core-collapse SNe powered by interaction between the ejecta and the dense circumstellar material. Among them, long-lived Type~IIn events are characterized by luminous, long-duration light curves with high radiative energy. Several cases of long-lived type IIn SNe exhibit substantial dust emission at late times. However, well-observed examples remain limited, and the details of their dust formation mechanisms remain poorly understood. Here we present photometric and spectroscopic observations of the Type~IIn SN~2024dy in ultraviolet, optical, and near-infrared (NIR) wavelength for $500$ days. SN~2024dy reached a peak magnitude of $M_r=-19.2$~mag with a total radiation energy of $1.9\times10^{50}$~erg. A NIR excess emerged at late phases, and the spectral energy distribution modeling indicates the presence of carbon dust with temperatures of $1300$-$1800$~K and masses of about $10^{-5}\ M_\odot$. The spectra features were typical of long-lived Type~IIn SNe. The late time H$\alpha$ profile exhibits a strong suppression of the red wing, providing evidence for newly formed dust. Our results suggest that the derived dust mass above may be underestimated due to optical depth effects. SN~2024dy provides an important observational case for understanding dust formation in Type~IIn SNe.

\end{abstract}

\keywords{Supernova: individual (SN 2024dy), Stellar mass loss, Circumstellar matter}

\section{Introduction} \label{sec:intro}
Massive stars with initial masses above $8$--$10\ M_\odot$ end their lives as core-collapse supernovae (SNe). Among them, those showing narrow hydrogen emission lines in their spectra are classified as Type~IIn, whose radiation is powered by shock interaction between the SN ejecta and the circumstellar material (CSM)~\citep{schlegel1990,filippenko1997,smith2017}. Type~IIn SNe exhibit diversity in peak luminosity and light-curve timescales, which is generally interpreted as reflecting differences in pre-explosion mass-loss histories of the progenitor~\citep[e.g.,][]{Kiewe2012,nyholm2020,hiramatsu2024}.

Among these events, the luminous and longest-lived examples are often referred to as “long-lived Type~IIn” SNe. These events are interpreted as explosions occurring within very massive CSM shells of a few $M_\odot$~\citep[][for a review]{fraser2020}. A prominent case is SN~2010jl, whose total radiated energy exceeds $10^{50}\,\mathrm{erg}$ \citep{fransson2014}, requiring an exceptionally large explosion energy even under the assumption of efficient ejecta--CSM energy conversion. Polarimetric observations further reveal pronounced asymmetry, suggesting a geometrically complex shock structure and CSM distribution \citep{patat2011}. Such asymmetry in the CSM has been proposed to significantly affect both the radiative efficiency and observed spectral properties of Type~IIn SNe~\citep{suzuki2019}.

Furthermore, long-lived Type~IIn SNe have emerged as key laboratories for studying dust formation in strongly interacting SN environments. Compared to the more common Type~IIP SNe, Type~IIn events more frequently exhibit luminous late-time near-infrared (NIR)/Mid-infrared (MIR) emission, often implying dust masses that are one to two orders of magnitude larger~\citep[e.g.,][]{fox2011,szalai2021}. Observationally, some events show pronounced NIR/MIR excesses accompanied by a progressive suppression of the red wing of H$\alpha$ on timescales of several hundred days (e.g., SN~2010jl;\citealt{fransson2014}). These features are commonly interpreted as signatures of newly formed dust that selectively obscures the receding side of the emitting region~\citep[e.g.,][]{chugai2018}.

However, the physical origin of the dust in Type IIn SNe remains debated. The observed blueshifted line profiles can be explained by continuous dust formation either in the SN ejecta or in the cool dense shell (CDS) formed between the forward and reverse shocks~\citep[e.g.,][]{smith2009,gall2014,bevan2019}. Alternatively, modeling of MIR spectral energy distributions suggests that a significant fraction of the emission may arise from pre-existing dust located at radii larger than the forward shock~\citep{fox2011}. In addition, multiple dust components may coexist, and the relative contributions are difficult to distinguish observationally.

Recent theoretical models predict that $10^{-3}$--$10^{-1}\, M_\odot$ of dust can form in the CDS of interacting SNe~\citep{sarangi2022}. This picture is also supported by long-term JWST/MIRI observations of Type~IIn SN~2005ip, which have revealed continued dust accumulation over more than a decade after the explosion \citep{shahbandeh2025}. Nevertheless, such estimates remain subject to significant uncertainties, particularly due to optical depth effects and the limited sensitivity of observations to cold dust at longer wavelengths.

In this context, adding to well-observed long-lived Type~IIn SN is valuable both as an addition to the still limited sample of this rare class and as a testbed for dust-formation physics in strongly interacting SNe. However, well-observed examples remain scarce, and the structure, temporal evolution, and optical depth evolution of the dust-forming regions are still poorly constrained. Thus, detailed multi-wavelength observations of new, well-characterized long-lived Type~IIn SNe are therefore essential for advancing our understanding of dust-formation mechanisms and interaction physics in this subclass.

In this work, we present $500$ days of ultraviolet (UV), optical, and NIR photometric and spectroscopic observations of SN~2024dy, a long-lived Type~IIn SN exhibiting clear signatures of dust formation. The dataset covers both the IR excess and the contemporaneous evolution of the H$\alpha$ line profile, allowing us to constrain the dust properties and give a stringent constraint on the dust mass. 
The structure of this paper is as follows. In Section~\ref{sec:obs and data}, we describe the observations and data reduction. Section~\ref{sec:photometry} presents the photometric results and Section~\ref{sec:spectra} provides the spectroscopic results. In Section~\ref{sec:analysis and discussion}, we discuss the energy budget and the properties, of the dust emission. A summary of this paper is given in Section~\ref{sec:summary}

\section{Observations and data reduction} \label{sec:obs and data}
\subsection{Discovery, classification and extinction} \label{ssec:discovery}
SN 2024dy (also known as GOTO24lk, PS24qe, ZTF24aaabljq, Gaia24acc, and ATLAS24ada) was discovered by \citet{tonry2024TNS} on  2 January 2024 (MJD 60311.5) with $18.60$ mag in orange band of the Asteroid Terrestrial-impact Last Alert System \citep[ATLAS;][]{tonry2018}.
The last non-detection was obtained by the Zwicky Transient Facility (ZTF) on 2023 December 29 (MJD $60307.48$), providing an upper limit of $\mathrm{m} \geq 16.87$ mag in the $g$-band. Just before this date, the Gravitational-wave Optical Transient Observer (GOTO) reported an upper limit of $\mathrm{m} \geq 19.8$ mag in the GOTO-$L$~band (MJD $60307.05$). Since the ZTF upper limit is relatively shallow, we adopt the GOTO upper limit as the final non-detection. In this paper, we take the midpoint between this GOTO non-detection and the first \textit{Gaia} detection as the explosion date (2023 December 31; MJD $60309.2$).

SN 2024dy was classified as a Type IIn SN at a redshift of $z = 0.0214$ by \cite{classification_wise2024}, using a spectrum obtained with the Lick-3m KAST spectrograph on 2024 January 8 (MJD 60317.55).
Since the host galaxy of SN~2024dy is faint and its distance has not been published, we adopt $z = 0.0214$ as the redshift, derived from the narrow H$\alpha$ emission in the spectrum used for classification. This corresponds to a distance of 90.1~Mpc and a distance modulus of $\mu = 34.77$ (adopting $H_{0} = 71~\mathrm{km~s^{-1}~Mpc^{-1}}$, $\Omega_{\mathrm{matter}} = 0.3$, and $\Omega_{\Lambda} = 0.7$, calculated with {\sc cosmotool}\footnote{\url{http://www.bo.astro.it/~cappi/cosmotools}}). Throughout this paper, we adopt this value.

We adopted the Galactic extinction toward SN 2024dy te be $E(B-V)_{\mathrm{MW}} = 0.026$ mag \citep{schlafly2011}. For the UV bands, we used the extinction law of \citet{Siegel2014} to estimate the extinction in each band. 
For the optical and near-infrared (NIR) extinction corrections, we used the extinction law of \citet{Cardelli1989} with $R_V = 3.1$. As for the host galaxy extinction, no Na {\sc i} D absorption was detected in any of our spectra (see Sect.~\ref{ssec:spec evo}); thus, no additional correction was applied.

\subsection{UV, optical and NIR photometry} \label{ssec:photo info}
The photometric data presented in this paper are a combination of public archive data and those obtained by our observations. 

Photometric observations in the $gi$-band were conducted over 53 epochs (between MJD~60351.66 and 60808.46) with the OPTCAM mounted on the 1-m telescope at Iriki Observatory, starting immediately after the discovery. The OPTCAM data were reduced in a standard manner, using the OKUDA pipeline\footnote{An image reduction Python-based pipeline developed for the data obtained with the 1-m telescope at Iriki Observatory}. The photometry was carried out with the point spread function (PSF) photometry using \texttt{IRAF}\footnote{IRAF stands for Image Reduction and Analysis Facility distributed by the National Optical Astronomy Observatory, operated by the Association of Universities for Research in Astronomy (AURA) under a cooperative agreement with the National Science Foundation.}/DAOPHOT package \citep{stetson1987}. The photometric calibration was performed using the field stars from the Pan-STARRS DR1 catalog \citep{chamber2016}.

NIR observations were conducted over 40 epochs (between MJD~60322.67 and 60807.55) using kSIRIUS \citep{nagayama2024} mounted on the 1.0-m telescope at the Iriki Observatory~\footnote{\url{https://agarc.sci.kagoshima-u.ac.jp/about/}} and HONIR \citep{akitaya2014} mounted on the 1.5-m Kanata telescope at the Higashi Hiroshima Observatory. The kSIRIUS data were analyzed in the same manner as those from OPTCAM; image reduction with the OKUDA pipeline followed by PSF photometry using DAOPHOT. The HONIR data were reduced using standard procedures in \texttt{IRAF}, and PSF photometry was then performed.
The calibration was performed using the stars from the 2MASS catalog \citep{skrutskie2006}.

For $g$ and $r$-band, we used data from the Zwicky Transient Facility \citep[ZTF;][]{Bell2019ZTF} obtained through the ALeRCE explorer\footnote{\url{https://alerce.science/}} \citep{foster2021}. Also, for the ATLAS $oc$-band, we used magnitude data obtained by using the ATLAS forced photometry server\footnote{\url{https://fallingstar-data.com/forcedphot/}} \citep{tonry2018,smith2020atlas,heinze2018atlas}. We binned the data into one-day intervals and adopted only the data with $\geq 5\sigma$ detections.

UV and $UBV$-band photometric data were obtained from publicly available data from the UVOT at the Neil Gehrels Swift Observatory \citep{roming2005}. Aperture photometry was performed using the UVOT data analysis software package HEAsoft\footnote{\url{https://heasarc.gsfc.nasa.gov/docs/software/heasoft/}}, following standard procedures.

\subsection{Optical spectroscopy} \label{ssec:spec info}
Spectroscopic observations were carried out using two facilities. We obtained 10 epochs of data with the Hanle Faint Object Spectrograph Camera (HFOSC) mounted on the 2.0 m Himalayan Chandra Telescope (HCT) of the Indian Institute of Astrophysics (IIA) over a period of approximately 400 days. The data were reduced using standard \texttt{IRAF} procedures.
We also obtained 4 epochs of spectroscopic observations of SN~2024dy with the ARIES-Devasthal Faint Object Spectrograph and Camera (ADFOSC) on the Devasthal Optical Telescope (DOT). We reduced the data using standard \texttt{IRAF} tasks, which include preprocessing, extraction of the 1D spectrum, wavelength calibration, and flux calibration.

In addition, two spectra available on Transient Name Server (TNS, $6$~d; Lick-$3$m/KAST and $7$~d; ESO-NTT/EFOSC2) are also included in our analysis.
The log of all spectroscopic observations is provided in table~\ref{tab:speclog}.

\section{Photometric properties} \label{sec:photometry}
\subsection{Light curves evolution} \label{ssec:lc evo}

\begin{figure*}[htbp]
    \centering
    \includegraphics[width=0.9\textwidth]{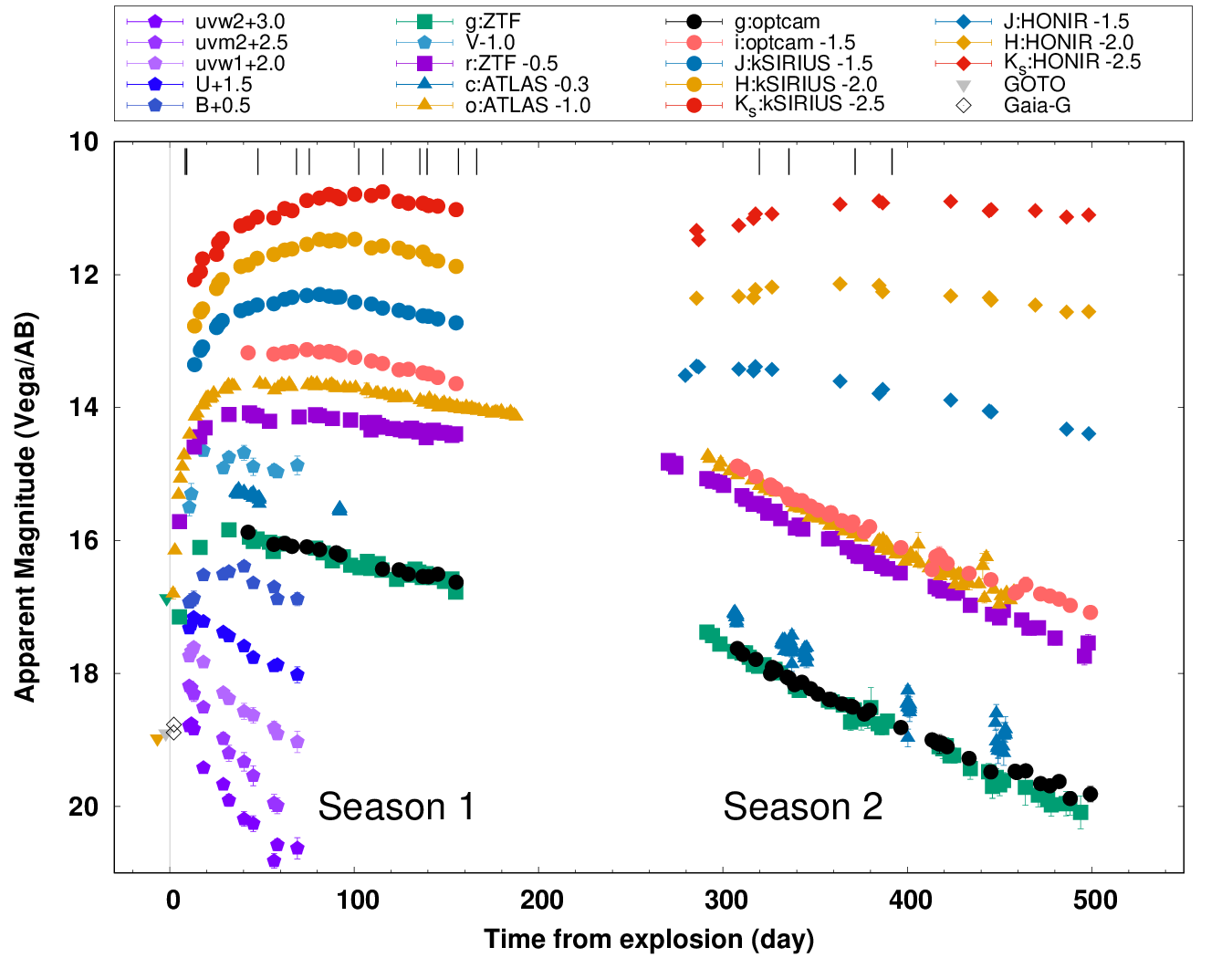}
    \caption{Multi-band light curves of SN 2024dy. 
   Each band is shifted in magnitude to distinguish them. 
   The black lines at the top of this panel present the spectroscopic epoch. The gray line denotes the estimated explosion date (MJD~60309.2, defined in Section~\ref{ssec:discovery}).}
    \label{multibandlc}
\end{figure*}

SN~2024dy is one of the long-lived Type~IIn SNe. 
Figure~\ref{multibandlc} shows the multi-band light curves of SN~2024dy. For the following discussion, we define the period up to MJD~$60497$ ($188$~d) as Season~1, and that after MJD~$60579$ ($270$~d) as Season~2.

In the UV bands, a rapid decline is observed from the first epoch, and the UV colors become progressively redder with time. The rise is well sampled in the optical $g$ through $i$-bands, with particularly high-cadence coverage in the $o$-band. 
Following \citet{ofek2014}, we model the early-time $o$-band light curve using a $t^{2}$ prescription given by

\begin{equation}
F = F_{\rm max}
\left(
1 - \left[ \frac{t - t_{\rm max}}{\Delta t} \right]^{2}
\right).
\label{eq:ofek_t2}
\end{equation}

Here, $F_{\rm max}$ is the peak luminosity, $t_{\rm max}$ is the time of maximum light derived from the parabolic fit, and $\Delta t$ is the time interval between zero luminosity and peak, corresponding to the rise time.
The fit yields a peak time of $t_{\max}=60347.54\pm0.15$ and a rise time of $\Delta t=38.03\pm0.17$ days. The corresponding explosion date inferred from these values is $t_{\rm exp}=60309.51\pm0.23$, which is broadly consistent with the explosion date adopted in this study (MJD~$60309.2$).

The rise time of SN~2024dy places it among the subclass of Type~IIn SNe with relatively long rise timescales identified in previous studies~\citep[see e.g.,][]{nyholm2020}. Around maximum light, the optical light curves show an almost flat evolution.
The time to reach peak brightness is longer in redder wavelengths, with the $K_s$-band exhibiting a rise time of about $90$ days.

The post-peak evolution is one of the most remarkable features of this object. In Season~1, after reaching peak, the $r$-band showed a gradual decline at a rate of $0.0036~\mathrm{mag\,d^{-1}}$. The decline rate is found to be larger at shorter wavelengths (e.g., decline rate for $g$-band is $0.0061~\mathrm{mag\,d^{-1}}$). In Season~2, when observations were resumed, the decline rate in the optical bands increased, while a rebrightening was observed in the NIR. During this period, the optical decline rate remained approximately constant, being $0.013$~mag~day$^{-1}$ in the $r$-band and $0.013$~mag~day$^{-1}$ in the $g$-band. 
In contrast, the NIR evolution was different: the $J$-band light curve exhibited a nearly flat evolution followed by a slow decline, whereas the $K_s$-band light curve showed a rebrightening around $t \sim 300-400$ d before transitioning to a nearly flat evolution.

\subsection{Light curve comparison} \label{ssec:phot compa}
\begin{figure}[htbp]
    \centering
    \includegraphics[width=0.5\textwidth]{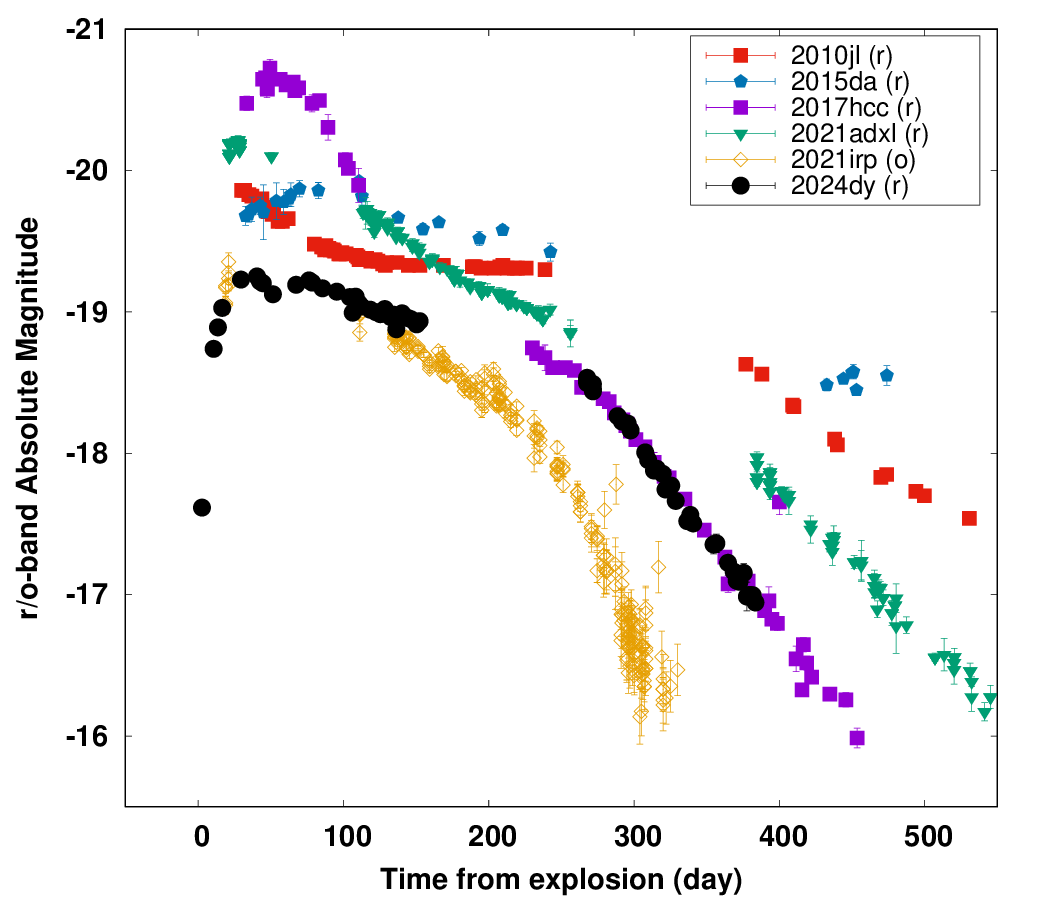}
    \caption{Comparison of the absolute-magnitude light curve of SN 2024dy with those of several long-lived Type IIn SNe:~SN 2010jl~\citep{fransson2014}, SN 2015da~\citep{tartaglia2020}, SN 2017hcc~\citep{moran2023}, SN 2021adxl~\citep{brennan2024}, and SN 2021irp~\citep[o-band;][]{reynolds2025}.
The explosion epochs and extinction corrections are adopted from the respective references.}
    \label{fig:optlc_compa}
\end{figure}

Figure~\ref{fig:optlc_compa} compares the $r$-band light curve of SN~2024dy
with the optical ($r$ and $o$-bands) light curves of several well-observed,
luminous, long-lived Type~IIn SNe, selected based on the
long timescales.
The comparison sample includes SN~2010jl~\citep{fransson2014},
SN~2015da~\citep{tartaglia2020},
SN~2017hcc~\citep{moran2023},
SN~2021adxl~\citep{brennan2024}, and
SN~2021irp~\citep{reynolds2025}. Although SN~2021irp is not formally classified as a Type~IIn SN, we include it here primarily because its light-curve morphology resembles that of SN~2024dy, with additional similarities in its late-time observational properties (further details are discussed in Section~\ref{ssec:spec compa}).

The optical rise time of SN~2024dy ($38.03\pm0.17$ d) is somewhat shorter than those reported for SN~2015da ($100\pm3$ d in the $R$ band;~\citealt{tartaglia2020}) and SN~2017hcc ($57\pm2$ d in the $o$ band;~\citealt{moran2023}). Nevertheless, within the broader population of Type~IIn SNe studied statistically, SN~2024dy would be classified as a subgroup with relatively long rise timescales~\citep[see e.g.,][]{nyholm2020}. For SN~2010jl, SN~2021adxl, and SN~2021irp, the rise to peak brightness was not sufficiently captured, and therefore their rise times are not tightly constrained~\citep[see][]{stoll2011,brennan2024,reynolds2025}. In Figure~\ref{fig:optlc_compa}, we adopt the explosion dates reported in the respective references for each object.

The comparison objects selected in this study are characterized by their luminosity and long-lived light-curve evolution. However, their peak luminosities and behavior near peak exhibit a large diversity. The peak absolute magnitude of SN~2024dy is $M_r = -19.2$~mag, which is roughly consistent with the median $r$-band peak magnitude estimated from recent statistical Type IIn samples (see \citealt{hiramatsu2024}). SN~2024dy is among the faintest in the comparison sample considered here, comparable to SN~2021irp at peak.

The decline rates among the comparison objects show notable similarities. SN~2010jl exhibits values of $\sim0.001$--$0.007$~mag~d$^{-1}$ between $100$ and $250$~d in $r$-band, and comparable decline rates are also seen in SN~2015da ($r$-band. $100$--$600$~d) and SN~2017hcc ($r$ band, $230$--$280$~d). These values are consistent with the decline rate of SN~2024dy in Season~1 ($r$-band, $0.0036$~mag~d$^{-1}$). All of these decline rates are smaller than the $^{56}$Co decay rate under the assumption of full gamma-ray trapping, as also mentioned by \citet{nagao2025}. An interesting feature is the difference in decline rates after $\sim250$~d. SNe~2010jl, 2017hcc, and 2021adxl show an accelerated optical decline similar to that of SN~2024dy. In contrast, SN~2015da does not exhibit a clear acceleration during the same period, though the data coverage is relatively sparse. Among the objects showing accelerated declines, there is a diversity in the decline rates: SN~2021irp shows the largest decline rate ($0.027$~mag~d$^{-1}$; \citealt{reynolds2025}), whereas SN~2010jl shows the smallest ($0.0049$~mag~d$^{-1}$; \citealt{fransson2014}).

\subsection{Blackbody fit and pseudo-bolometric luminosity} \label{ssec:bbfit}

\begin{figure}[htbp]
    \centering
    \includegraphics[width=0.45\textwidth]{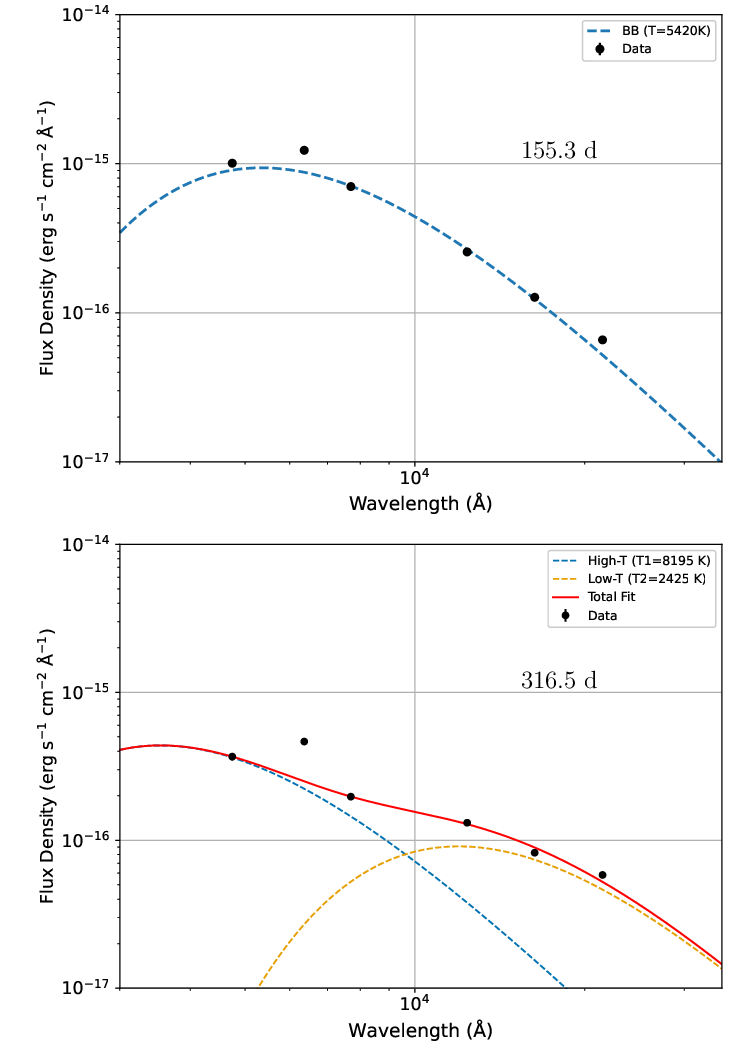}
    \caption{Examples of the SED fitting with BB models. The upper panel shows a single-BB fit to a Season~1 (155.3d) SED, and the lower panel shows a two-BB fit to a Season~2 (316.5d) SED. The observed SEDs are shown with filled circles. In the lower panel, the dashed lines represent the hot and cool BB components, and the red solid line shows their sum. The cool BB component accounts for the NIR excess observed in Season~2.}
    \label{fig:SEDfit}
\end{figure}

\begin{figure}[htbp]
    \centering
    \includegraphics[width=0.5\textwidth]{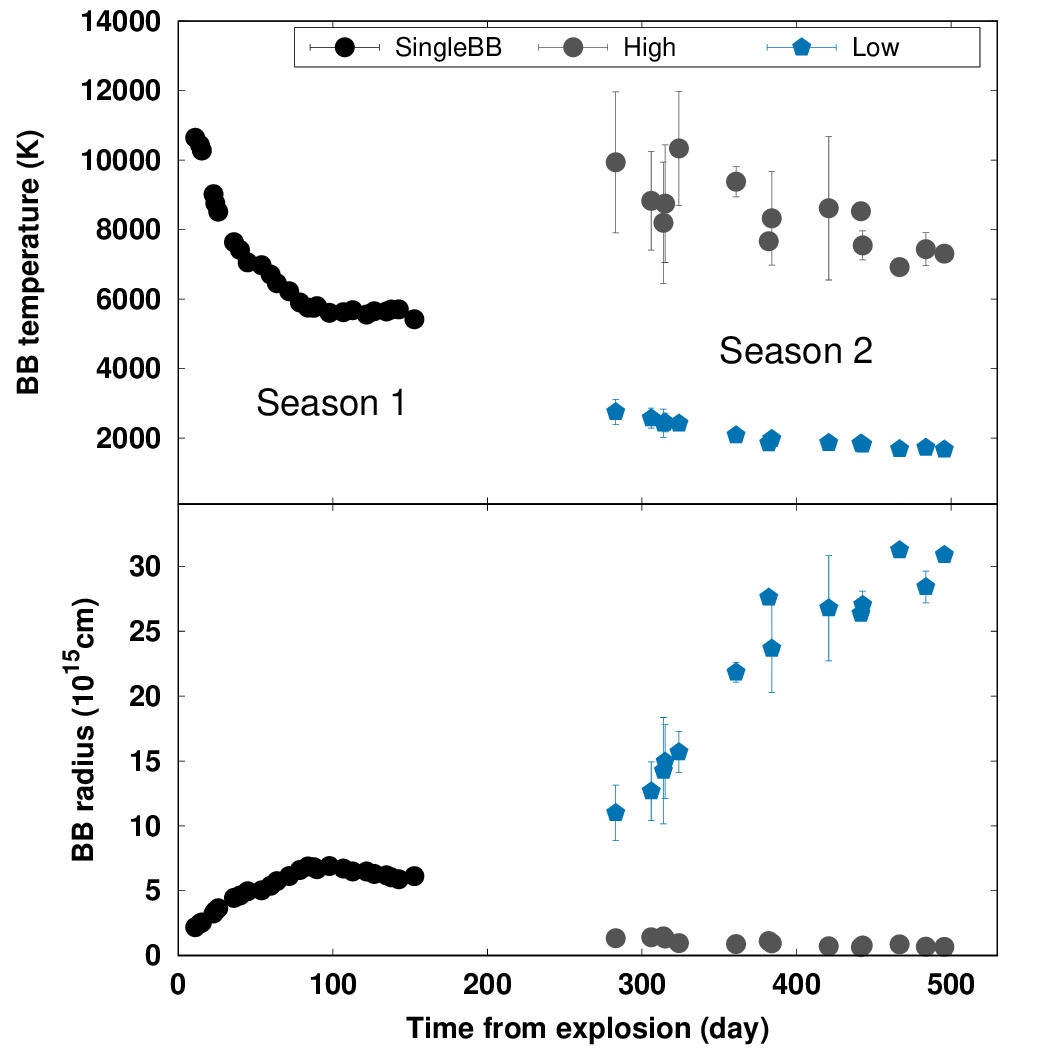}
    \caption{Temporal evolution of the temperature (top) and radius (bottom) obtained from the blackbody fits. Black, gray, and blue points represent the single-component, hot-component, and cool-component fits, respectively.}
    \label{bbfit}
\end{figure}

To investigate the NIR rebrightening observed in Season~2, we performed a two-component blackbody (BB) fit to the SED. Figure~\ref{fig:SEDfit} shows examples of the BB fitting: a single-BB model for a Season~1 SED and a two-BB model for a Season~2 SED. The two-BB model is used as a phenomenological description to estimate the temperatures and BB radii of the optical and NIR components. The fit reveals a clear IR excess in Season~2, indicating the presence of a low-temperature component with $T \lesssim 3000$~K. The results of the two-component BB fit are presented in Figure~\ref{bbfit}.

In Season~1, the SED was well reproduced by a single hot BB component with a temperature exceeding $10,000~\mathrm{K}$ at early times. The temperature decreased from $\sim11{,}000~\mathrm{K}$ to $<6000~\mathrm{K}$ within the first $\sim100$~days after the explosion. The inferred BB radius increased from $\sim2.2\times10^{15}$~cm to $\sim6.9\times10^{15}$~cm in this period. Subsequently, the temperature became nearly constant, and the radius gradually decreased.

In Season~2, the observed SEDs were well reproduced by a two-component model consisting of a hot component dominating the optical emission ($\sim8000~\mathrm{K}$) and a cool component dominating the NIR emission ($<3000~\mathrm{K}$). 
The apparent increase in the temperature of the hot component
between Seasons~1 and 2 should be interpreted with caution because the optical
SEDs in Season~2 are constrained by fewer bands, resulting in relatively large
uncertainties. A similar increase in the hot-component temperature was reported
for SN~2010jl by \citet{fransson2014}, but its physical origin is not uniquely
constrained by the present data.
The temperature of the cool component decreased from approximately $3000~\mathrm{K}$ to $1800~\mathrm{K}$, while the radius increased from $\sim1.1\times10^{16}~\mathrm{cm}$ to $\sim3.1\times10^{16}~\mathrm{cm}$. The origin of this cool component would be newly formed dust: more detail will be discussed in Section~\ref{sssec:dustmodel}.

\begin{figure}[htbp]
    \centering
    \includegraphics[width=0.5\textwidth]{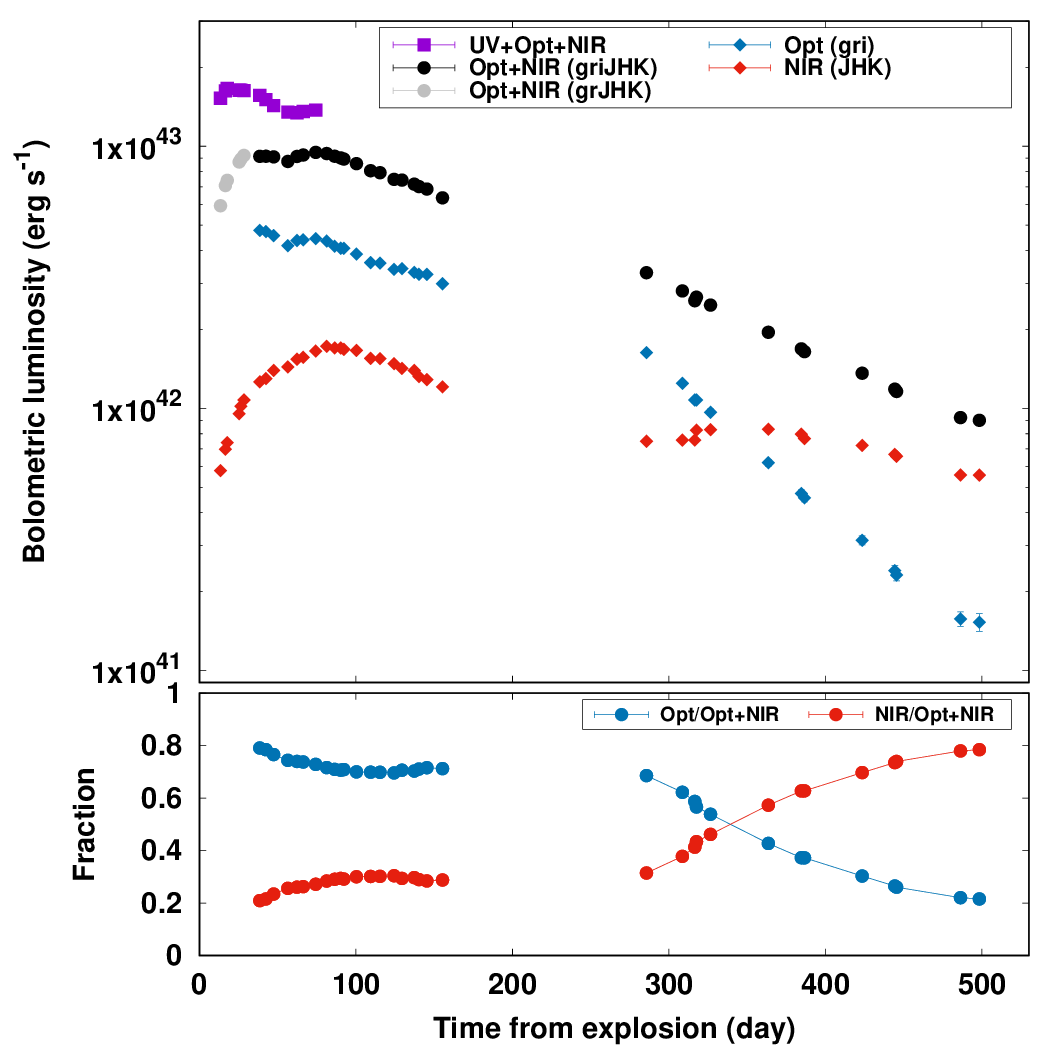}
    \caption{Upper panel: Pseudo-bolometric light curve derived from the optical--NIR SEDs (black) and from the UV--optical--NIR SEDs (purple). The optical and NIR luminosities are also shown (blue and red, respectively). Epochs at early phases for which $i$-band data are unavailable are shown in gray. Lower panel: Temporal evolution of the optical ($gri$) and NIR ($JHK_s$) contributions to the bolometric luminosity ($griJHK_s$).}
    \label{fig:lbol_fraction}
\end{figure}

We also derived the pseudo-bolometric luminosity by integrating the SEDs over wavelength using the trapezoidal assumption. To ensure consistency across all epochs, the pseudo-bolometric luminosity was constructed using only the bands with coverage from early to late phases. The resulting pseudo-bolometric light curve is shown in Figure~\ref{fig:lbol_fraction} (black points). At early epochs where $i$-band data are unavailable, the pseudo-bolometric luminosities at these epochs are shown in gray. Because UV observations were obtained only at early phases, the UV--optical--NIR pseudo-bolometric light curve is shown only for the early epochs (purple). For comparison, we also plot the optical luminosity (blue points), obtained by integrating the flux in the $gri$-bands, and the NIR luminosity (red points), obtained from the $JHK_s$-bands. The lower panel of Figure~\ref{fig:lbol_fraction} shows the temporal evolution of the optical and NIR fractions of the bolometric luminosity.

During Season~1, the UV--optical luminosity dominates and accounts for the majority of the total radiated energy. The peak luminosity of the UV--optical--NIR pseudo-bolometric light curve is estimated to be $\sim1.6\times10^{43}\,\mathrm{erg\,s^{-1}}$ and the optical--NIR pseudo-bolometric light curve reaches a peak luminosity of $9.5\times10^{42}\,\mathrm{erg\,s^{-1}}$. 
The total radiated energy in the optical--NIR range inferred from the pseudo-bolometric light curve up to the end of Season~2 is estimated to be $1.9 \times10^{50}$~erg.

From Season~2, the optical luminosity exhibits a rapid decline, whereas the NIR luminosity evolves more steadily. As a result, the relative contributions of the optical and NIR emission become reversed. The NIR fraction reaches $\sim60\%$ at $\sim400~\mathrm{d}$ after the explosion and increases to $\sim80\%$ around $500~\mathrm{d}$  (shown in lower panel of figure~\ref{fig:lbol_fraction}).

\section{Spectroscopic properties} \label{sec:spectra}
\subsection{Spectroscopic evolution} \label{ssec:spec evo}

\begin{figure*}[htbp]
    \centering
    \includegraphics[width=0.9\textwidth]{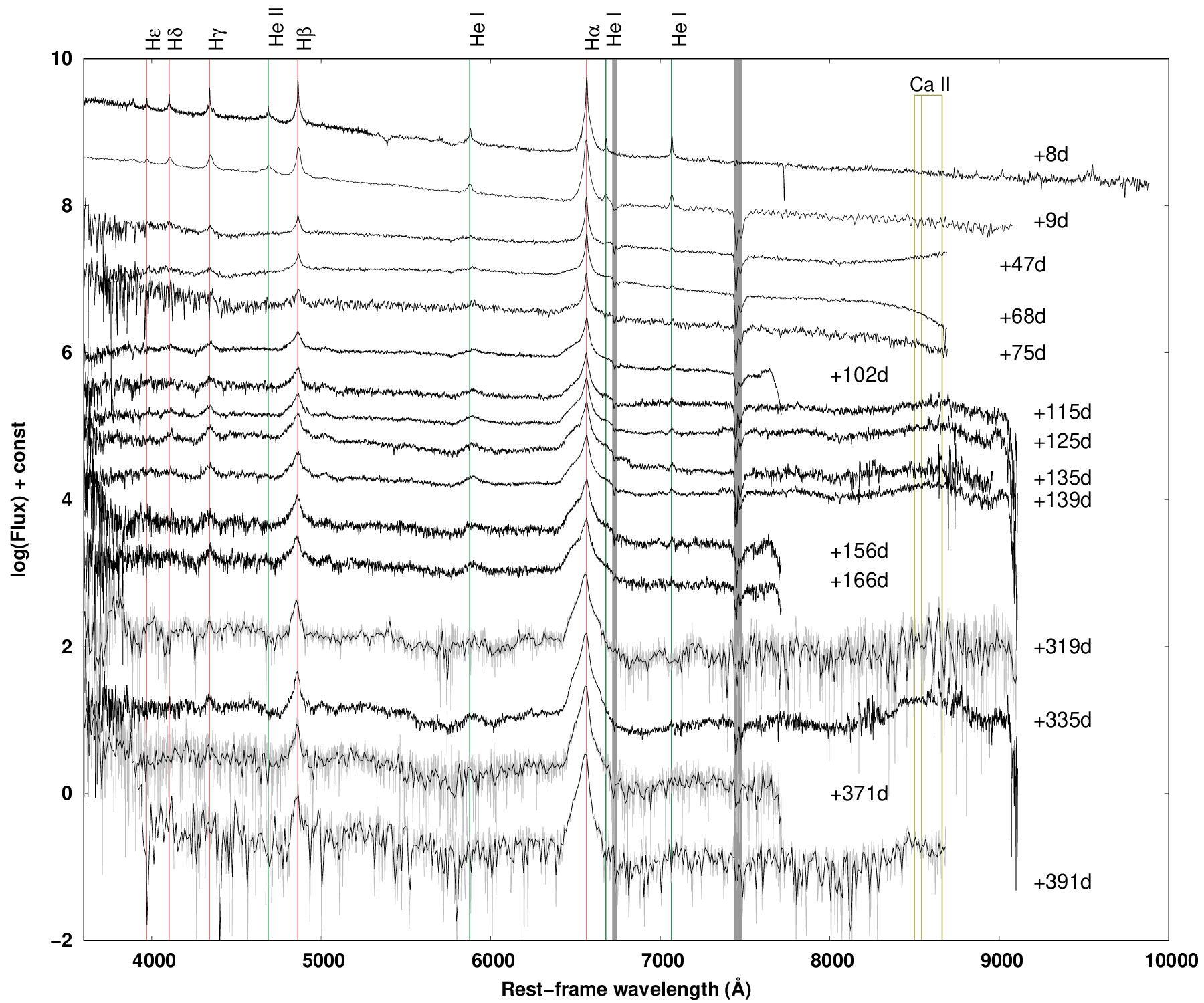}
    \caption{The spectra of SN~2024dy. The first two (upper side) spectra were published on TNS. The numbers indicate the days since the estimated explosion date. Several spectra obtained during Season~2 have been smoothed for clarity. The grey lines around $6800$ and $7500$~\AA\ indicate the telluric absorption features, which are corrected in some spectra.}
    \label{fig:allspectra}
\end{figure*}

Figure~\ref{fig:allspectra} shows all the spectra of SN~2024dy we obtained. Two spectra obtained by other groups soon after the discovery (available on TNS) are also shown at the top of Figure~\ref{fig:allspectra} ($8~\mathrm{d}$ and $9~\mathrm{d}$). Several late-time spectra have been smoothed for clarity.

As in typical Type~IIn SNe observed the early spectra, SN~2024dy is characterized by a blue continuum together with prominent narrow (+intermediate) Balmer emission lines (H$\alpha$ $\lambda6563$, H$\beta$ $\lambda4861$, H$\gamma$ $\lambda4340$, H$\delta$ $\lambda4101$, and H$\epsilon$ $\lambda3970$). Narrow He~{\sc i} $\lambda\lambda5875,7065$ and He~{\sc ii} $\lambda4685$ emission lines are also present shortly after the discovery. 
Around the He~{\sc ii} $\lambda4686$ line, a somewhat broad emission component is detected only in the earliest spectrum. This feature is likely to be C~{\sc iii} $\lambda\lambda4647,4650$ and/or N~{\sc iii} $\lambda\lambda4634,4642$ lines that are observed in very early spectra of Type~II SNe (see, e.g., \citealt{khazov2016}; \citealt{yaron2017}). Unfortunately, there is a temporal gap between the first two spectra and the third spectrum. The third spectrum ($47$~d after the explosion) shows a very blue continuum. The He emission lines disappear and only the Balmer-series lines remain. In the spectra obtained after $102$~d, a shallow absorption feature appears on the blue side of He~{\sc i} $\lambda5876$. Also, the H$\alpha$ profile evolves from the early narrow and intermediate components into a profile that includes a broad emission component. The evolution of the H$\alpha$ profile is discussed in section~\ref{ssec:line profile}.

In Season~2 ($>270$ d), four spectra were obtained. Some of the spectra have low S/N, and smoothing has been applied to improve their visibility. Except for the $335$~d spectrum, the data are noisy, making it difficult to clearly identify features other than H$\alpha$ and H$\beta$. Nevertheless, all spectra show a broad component in H$\alpha$, accompanied by an eroded red wing. In the highest-quality spectrum at $335$~d obtained with the HCT, a shallow and extended absorption feature on the blue side of He~{\sc i} $\lambda5875$ can be seen in addition to H$\alpha$ and H$\beta$. We also detect broad emission lines around $8500$~\AA\ appear to correspond to the Ca~{\sc ii} NIR triplet ($\lambda\lambda8498,8542,8662$).

\subsection{Spectroscopic comparison} \label{ssec:spec compa}
\begin{figure*}[htbp]
    \centering
    \includegraphics[width=0.85\textwidth]{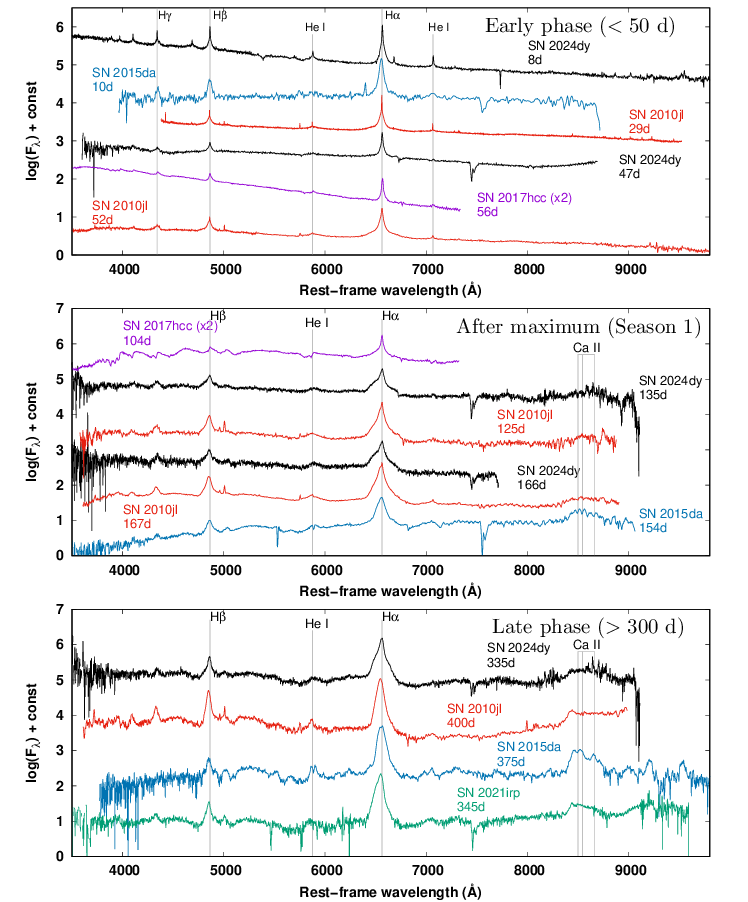}
    \caption{Comparison between the spectra of SN~2024dy and those of other long-lived Type~IIn SNe. Top panel: early phase ($<50$~d). Middle panel: post-peak phase ($50$~d to the end of Season~1). Bottom panel: late phase ($>300$~d). The redshifts for SN~2010jl, SN~2015da, SN~2017hcc and SN~2021irp were adopted from the literature cited below; \citet{fransson2014,tartaglia2020,moran2023,reynolds2025}. For visual clarity, the spectrum of SN~2017hcc was plotted as $2\log(F)+{\rm const}$ and is labeled as "SN~2017hcc ($\times 2$)".}
    \label{fig:spec_compa_all}
\end{figure*}

In Figure~\ref{fig:spec_compa_all}, we compare representative spectra of SN~2024dy at different evolutionary phases with those of other long-lived Type~IIn SNe. The comparison spectra were obtained from WISeREP, and the adopted redshift values follow those reported in the respective references given in the caption. 

In the early phase ($<50$ d), SN~2024dy shows a significant similarity to SN~2010jl. The H$\alpha$ emission line profiles in both objects are nearly identical, and both display wings that resemble symmetric Lorentzian. In addition to the H$\alpha$, narrow He~{\sc i} and He~{\sc ii} emission lines were also observed.
Apart from the Balmer series, no prominent additional spectral features are detected at this stage.

During the intermediate phase ($50$-$170$ d), more spectral features of SN~2024dy start to develop. As shown in Figure~\ref{fig:spec_compa_all}, SN~2024dy, SN~2010jl, and SN~2015da all develop asymmetric H$\alpha$ profiles, in contrast to the more symmetric shapes in the earlier phase. In this phase, the red side of the profile becomes increasingly suppressed, while the blue side remains extended. By contrast, SN~2017hcc shows a symmetric Lorentzian-like H$\alpha$ profile even at comparable epochs, highlighting the diversity of line profile evolution among long-lived Type~IIn SNe.

In the late phase ($>300$ d), the H$\alpha$ profile of SN~2024dy evolves further showing a close resemblance to that of SN~2021irp. Similarities are also found with the late-time spectral evolution of SN~2021acya, as discussed by \citet{salmaso2025}, particularly in the pronounced asymmetry of the Balmer-line profiles. A common property shared by SN~2024dy and all comparison objects is the persistent absence of nebular emission features typically observed in late-time spectra of ordinary Type~II SNe (such as the [O~{\sc i}] $\lambda\lambda6300,6364$ doublet) even at later epochs.

\subsection{Line profile evolution} \label{ssec:line profile}

\begin{figure}[htbp]
    \centering
    \includegraphics[width=0.45\textwidth]{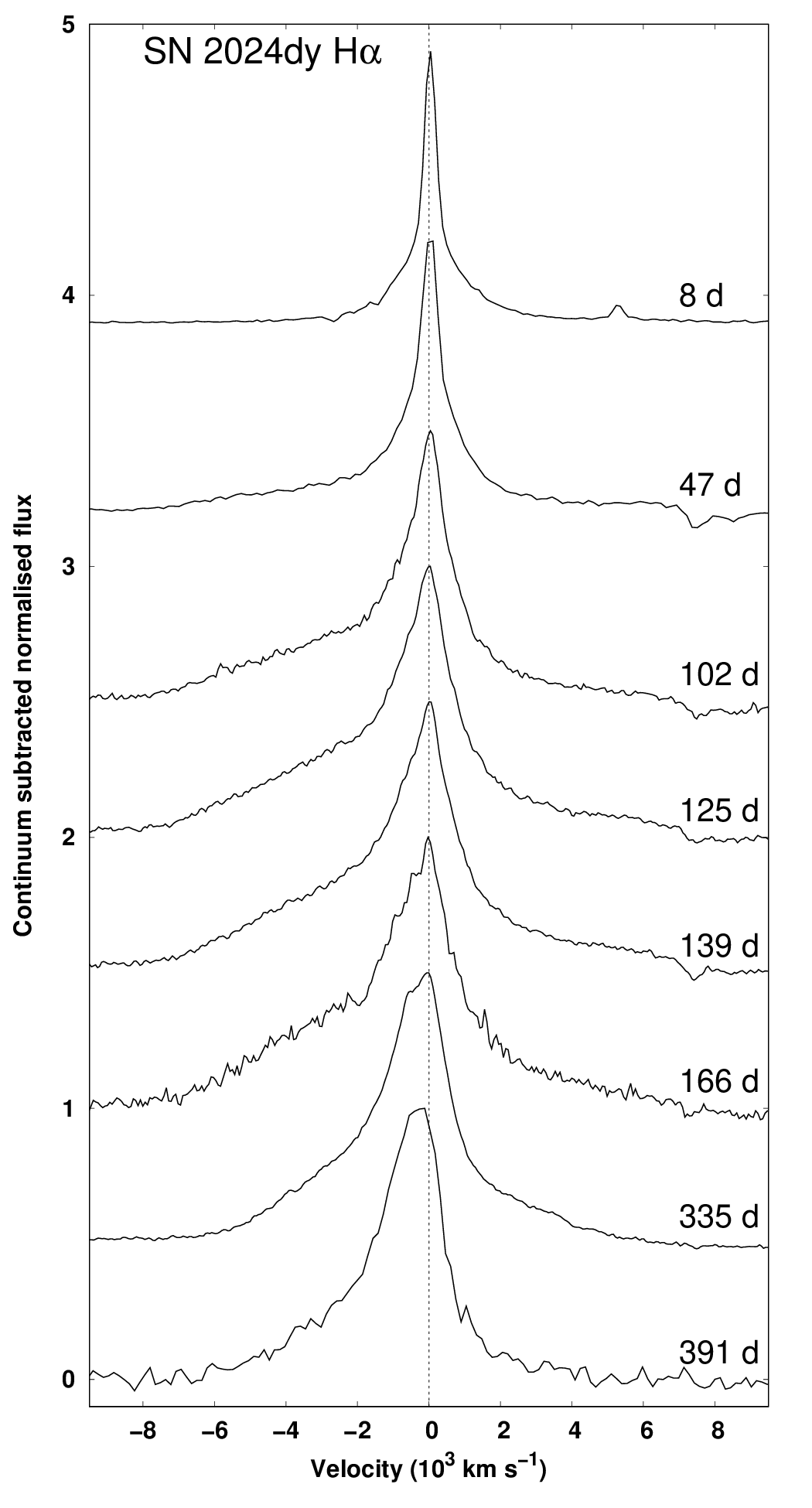}
    \caption{Evolution of the H$\alpha$ emission line. The continuum has been subtracted, and each profile is normalized to the peak of the narrow component.}
    \label{fig:ha_norma}
\end{figure}

\begin{figure*}[htbp]
    \centering
    \includegraphics[width=0.85\textwidth]{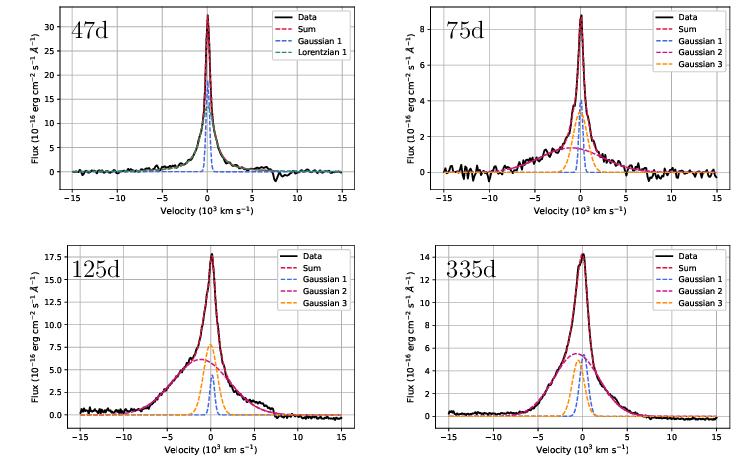}
    \caption{Fits to the H$\alpha$ line profiles at $47$, $75$, $135$, and $335$~d. The $47$~d profile is modeled with a combination of a Gaussian and a Lorentzian component, while the later epochs are fitted using multiple Gaussian components.}
    \label{fig:hafit_result}
\end{figure*}

One of the most notable features of SN~2024dy is the evolution of its H$\alpha$ profile. Figure~\ref{fig:ha_norma} shows the expanded H$\alpha$ profiles, and Figure~\ref{fig:hafit_result} presents the results of multi-component Gaussian or Lorentzian fits for the spectra at several epochs. The fitting results for all the epochs are summarized in Table~\ref{tab:ha_profile}.

The evolution of the H$\alpha$ line can be divided into three distinct phases. In the early phase ($8\sim47$~d), the profile consists of a combination of narrow Gaussian and intermediate-width Lorentzian. Such emission line profiles are commonly observed in the early spectra of Type~IIn SNe (e.g., 2010jl;~\citealt{fransson2014}). The intermediate width component shows a symmetric profile, with a Lorentzian FWHM of $1500\sim1800$~km~s$^{-1}$. 

The narrow component has a FWHM of $\sim300$~km~s$^{-1}$ in the first spectrum, which has the highest spectral resolution among our spectra. The FWHM of the narrow Gaussian component measured in the $8$~d spectrum is $\sim300~{\rm km~s^{-1}}$. However, the spectral resolution of the Lick/Kast spectrum is estimated to be $\sim390~{\rm km~s^{-1}}$ around H$\alpha$. Therefore, this narrow component is not spectrally resolved, and the measured width should be regarded as an upper limit on the velocity of the unshocked CSM. The larger values measured at $9$~d and $47$~d are also likely affected, or dominated, by the instrumental resolution at those epochs.

The second phase corresponds to the period from the peak to the end of Season~1 ($68$--$166$~d). During this phase, the intermediate-width component is better reproduced by a Gaussian profile, and a blueshifted broad component begins to emerge. The FWHM of this broad component is $8000$--$7000$~km~s$^{-1}$ and shows little evolution through the end of Season~1. The peak of the broad component also remains nearly constant at $\sim-1000$~km~s$^{-1}$. 

In the late phase from Season~2 onward, the H$\alpha$ profiles consistently exhibit a well-developed broad component and show a characteristic asymmetry caused by the suppression of the red wing. This line asymmetry becomes more apparent when comparing the blue and red sides of the profile, as shown in Figure~\ref{fig:ha_reverse}, where the blue side has been reflected across the rest wavelength. In the $335$~d spectrum, which has the highest S/N in Season~2, a similar shape may be present in the H$\beta$ profile. 

\begin{figure}[htbp]
    \centering
    \includegraphics[width=0.5\textwidth]{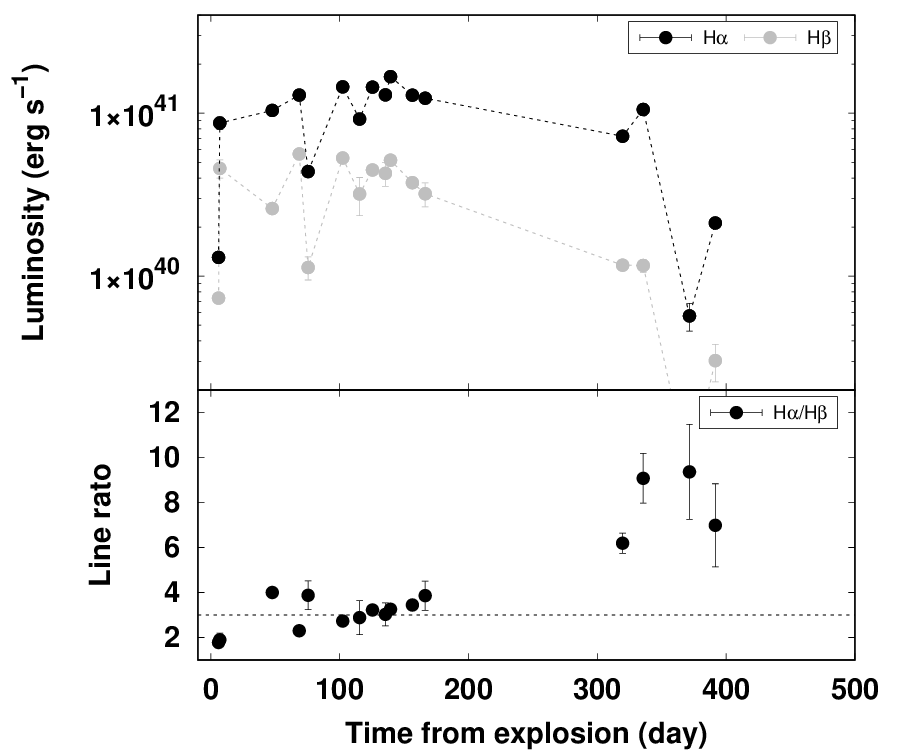}
    \caption{Upper: Evolution of the H$\alpha$ (black) and H$\beta$ (gray) line luminosities. Lower: Temporal evolution of the H$\alpha$/H$\beta$ flux ratio. The expected Case~B recombination value (H$\alpha$/H$\beta \approx 3$) is indicated by the dashed line.}
    \label{fig:line_ratio}
\end{figure}

Figure~\ref{fig:line_ratio} shows the temporal evolution of the H$\alpha$ and H$\beta$ luminosities, as well as the H$\alpha$/H$\beta$ ratio. 
The bottom panel indicates the expected Case~B hydrogen recombination value (H$\alpha$/H$\beta \approx 3$; \citealt{osterbrock2006}) with a black dashed line. During Season~1, the ratio remains close to $\sim3$, whereas in Season~2 it increases to significantly higher values.

\section{Discussion} \label{sec:analysis and discussion}

\subsection{Energy source} \label{ssec:enegy source}
The Season~1 spectra of SN~2024dy are dominated by narrow hydrogen emission lines, indicating interaction between the H-rich CSM and the SN ejecta (see Section~\ref{sec:spectra}). The light curves exhibit a relatively slow rise ($\sim38$~d), which is also consistent with interaction with a dense CSM. The timescale can be explained by the photon diffusion timescale through an optically thick CSM. In addition, the observed timescales and peak luminosity clearly place SN~2024dy within the subclass of long-lived Type~IIn SNe.

Assuming that luminosity of the ejecta-CSM interaction is fed by energy at the shock front, the progenitor's mass-loss rate can be estimated using the relation given by \citet{chugai1994}:
\begin{equation}
\dot{M}=\frac{2L}{\epsilon}\frac{v_w}{v_{SN}^3}
\label{equ1}
\end{equation}
Here, $\dot{M}$ denotes the mass-loss rate of the progenitor star, and $\epsilon$ represents the efficiency with which the kinetic energy of the shock is converted into radiative energy. Following previous studies, we adopt a value of $\epsilon = 0.3$ in this work \citep[e.g.,][]{moriya2013}.
For the wind velocity of the CSM $v_w$, we adopt a steady wind velocity typical of LBVs, $v_w\sim100\ \mathrm{km\ s^{-1}}$~\citep{smith2017}.
For the velocity of the SN ejecta $v_{\mathrm{SN}}$, for which we adopt an expansion velocity of $v_{\mathrm{SN}}=7700$~km~s$^{-1}$ inferred from the temporal evolution of the BB radius. The peak luminosity is estimated to be $1.6\times10^{43}$~erg~s$^{-1}$ from the pseudo-bolometric light curve obtained by integrating the SED from the UV, optical~($g,r,i$) and NIR~($J,H,K_s$) (see section~\ref{ssec:bbfit}). 

From these values, the mass-loss rate is estimated to be $0.037~M_{\odot}~\mathrm{yr^{-1}}$. This value is consistent in order of magnitude with the mass-loss rates inferred for other Type~IIn SNe ($10^{-3}$–$10^{-1}\ M_{\odot}\,\mathrm{yr^{-1}}$ for some interacting SNe; \citealt{salmaso2025}). Although this value represents a lower limit, the mass-loss rate estimated for SN~2024dy is relatively small compared to those of the comparison objects. For SN~2010jl, a mass loss rate of $\gtrsim0.1\ M_{\odot}\,\mathrm{yr^{-1}}$ has been inferred \citep{fransson2014}, which is an order of magnitude higher than our estimate for SN~2024dy. Similarly, a mass-loss rate for SN~2015da is estimated to be $0.6$--$0.7\,M_{\odot}\,\mathrm{yr^{-1}}$ \citep{tartaglia2020}, and that for SN~2017hcc is $0.12\,M_{\odot}\,\mathrm{yr^{-1}}$ \citep{kumar2019}, both of which are substantially larger than that of SN~2024dy. Since the luminosity produced by CSM interaction is approximately given by Eqation~(\ref{equ1})~\citep[e.g.,][]{moriya2013}, it is natural that, for SN~2024dy, whose peak radiated energy is lower by an order of magnitude, yield a relatively modest mass-loss rate of $\dot{M} \sim 10^{-2}\ M_{\odot}\,{\rm yr^{-1}}$ is inferred. Nevertheless, this value still falls within the regime of a dense CSM.

It should be noted that the mass-loss rate derived above is subject to uncertainty, particularly due to assumptions about the geometry of the CSM. Luminous Type~IIn events such as SN~2015da \citep{tartaglia2020,smith2024} require very high mass-loss rates, corresponding to several solar masses of CSM when interpreted under the assumption of spherical symmetry. However, SN~2015da also shows evidence for pronounced asymmetry \citep{smith2024}, indicating that the true geometry of the CSM remains poorly constrained. In this context, SN~2024dy exhibits a shorter rise time than SN~2015da, which may reflect differences in the line-of-sight distribution of the CSM. Nevertheless, the present observations do not provide meaningful constraints on the degree of CSM asymmetry for either event. We therefore assume a spherically symmetric CSM distribution for simplicity.

An evaluation of the energy budget indicates that SN~2024dy radiated a total energy of $\sim2\times10^{50}$~erg. This value is of the same order of magnitude as the total radiated energy of $5.8\times10^{50}$~erg estimated for SN~2010jl \citep{fransson2014}. The total radiated energy from CSM interaction correlates with the total mass of the CSM swept up by the shock~\citep{moriya2013}. Therefore, if the luminosity is indeed dominated by CSM interaction, a comparable total radiated energy suggests a similar CSM mass. 
In this context, the combination of a comparable total radiated energy and a lower peak luminosity in SN~2024dy favors a scenario of ``moderate interaction,'' in which the CSM has a lower density but is distributed over a more spatially extended structure, resulting in a total CSM mass consistent with that inferred for SN~2010jl (several $M_{\odot}$).

\subsection{Origin of the line profile evolution} \label{ssec:line evolution}
In Type~IIn SNe and other strongly interacting SNe, many studies have reported that emission line profiles evolve toward systematically blueshifted and asymmetric shapes. In many cases, the early-time line profiles are nearly symmetric and characterized by Lorentzian wings produced by electron scattering. Pronounced blueshifts develop at later epochs, particularly in the intermediate-width component originating from post-shock CDS (e.g., SN~2010jl and SN~2015da; \citealt{smith2012,tartaglia2020,smith2024}). Such line profile asymmetries do not necessarily reflect intrinsic geometric asymmetries in the CSM, but may instead arise from selective suppression of emission from the receding side. Two main mechanisms are often invoked to explain this behavior (although they are not mutually exclusive). 
One is geometric obscuration, in which the visibility of the emitting regions depends on the viewing angle, and the interaction region and/or the unshocked CSM is optically thick. The other is obscuration by newly formed dust within the SN ejecta or the CDS, which selectively absorbs the redshifted emission. The latter dust formation scenario is strongly supported by its association with independent observational signatures, such as a rapid decline in optical luminosity and the emergence of an IR excess, as exemplified by events such as SN~2006jc (e.g., \citealt{mattila2008}).

The line profile asymmetry observed in SN~2024dy during the phase after its peak ($68$--$166$ d, as shown in Figures~\ref{fig:ha_norma} and~\ref{fig:hafit_result}) is not consistent with the canonical signatures of dust formation. During this epoch, the blue side of the H$\alpha$ emission-line profile extends to velocities of up to $\sim 8000\ \mathrm{km\ s^{-1}}$, whereas the red side is clearly more limited in velocity extent. However, neither accelerated optical fading nor the accompanying NIR excess is observed. The absence of these observational signatures suggests that dust is not the dominant cause of the line profile at this phase. 

Assuming that the H$\alpha$-emitting region is distributed in a shell-like structure and that an optically thick region (e.g., the interaction region or dense CSM) exists interior to it, the asymmetric line profile observed in SN~2024dy at this phase can be naturally explained. 
In this scenario, emission from the near side of the shell is observed with little attenuation, whereas emission from the far side is obscured by the optically thick interior region, resulting in a systematic suppression of the redshifted (receding) component. Consequently, an asymmetric profile characterized by a selective deficit on the red side is produced. 

In contrast, the blueshifted Lorentzian profile formed via electron scattering in a radiatively accelerated CSM, as discussed for SN~2010jl \citep{fransson2014}, produces a largely symmetric broadening with a shifted peak. Such a profile cannot account for the one-sided suppression of the red wing observed in SN~2024dy. Consistently, the observed line profile is better reproduced by a Gaussian component than by an intermediate-width Lorentzian (see Section~\ref{ssec:line profile}). Therefore, the line asymmetry in this object is more plausibly attributed to geometric obscuration by optically thick material rather than to a blueshift caused by radiative acceleration.

In Season~2, the asymmetric line profile is likely caused by dust absorption.
The NIR excess detected in SN~2024dy emerges at the beginning of Season~2,
coincident with the epoch at which the asymmetry in the H$\alpha$ profile becomes pronounced. The temporal correspondence between these phenomena strongly suggests that the H$\alpha$ profile in Season~2 results from the selective obscuration of the receding emission by newly formed dust. We therefore here examine whether the observed line profile can be explained by dust absorption, and defer further discussion of the dust origin and mass to Sections~\ref{ssec:dust origin} and~\ref{ssec:estimation dustmass}.

\begin{figure}[htbp]
    \centering
    \includegraphics[width=0.5\textwidth]{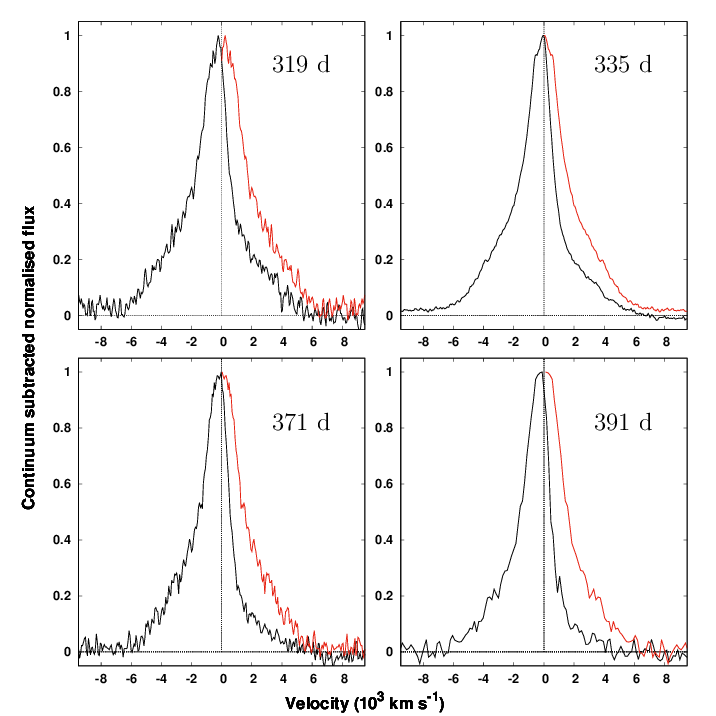}
    \caption{H$\alpha$ profiles in Season~2, with the blue side reflected onto the red side across the rest wavelength. The reflected profiles are shown in red, and the rest wavelength is indicated by a black dashed line.}
    \label{fig:ha_reverse}
\end{figure}

Figure~\ref{fig:ha_reverse} presents an enlarged view of the H$\alpha$ line profiles observed during Season~2, in which the blue side of the profile relative to the rest wavelength is reflected and overplotted onto the red side (red line). The asymmetry at this phase is characterized by an apparent attenuation affecting only the red side of the profile, with little suppression near the rest wavelength with no pronounced blueshift of the line peak. Using the reflected blue side profile as a reference, we estimated the fraction of flux lost on the red side. Specifically, for the red-side profile shown in Figure~\ref{fig:ha_reverse}, we integrated the flux over the velocity range of $0-5000\ \mathrm{km\ s^{-1}}$ and computed the ratio relative to the corresponding integrated flux on the blue side. For the final spectrum obtained at $391$ d, we find that the integrated flux on the red side is approximately $50\%$ of that on the blue side. This ratio remains nearly constant throughout Season~2, showing no significant temporal evolution at all epochs in Season~2.

\subsection{Dust as the origin of IR excess} \label{ssec:dust origin}

An IR excess observed from the NIR to MIR is one of the characteristic features of Type~IIn SNe. Two primary origins have been widely discussed: (1) heating of pre-existing CS dust by the SN radiation (IR echo), and (2) newly formed dust that condenses in the SN environment after explosion. If the dust is newly formed, it would condense either within the SN ejecta \citep[e.g.,][]{kozasa1989, wooden1993} or in the CDS in post-shock gas lying in between the forward and
reverse shocks~\citep[e.g.,][]{sarangi2022, smith2024}. 
From a large Spitzer sample of Type~IIn SNe, \citet{fox2011} found that many objects exhibit a late-time IR excess, which is interpreted as emission from pre-existing CS dust heated by the SN. In contrast, for long-lived Type IIn events such as SN~2010jl, SN~2005ip, and SN~2015da, dust formation in the CDS on timescales of several hundred days after explosion has been suggested, producing significant observational signatures including enhanced IR emission and asymmetric emission line profiles \citep[e.g.,][]{gall2014, sarangi2018, smith2024}. In the following section, we discuss the origin of the IR excess observed in SN~2024dy from two perspectives: an IR echo and dust formation.

\subsubsection{IR echo} \label{sssec:IR echo} 
First, we discuss whether the IR excess observed during Season~2 in SN~2024dy can be attributed to an IR echo. The BB radius of the cool component estimated in Section~\ref{ssec:bbfit} is $\sim10^{16}$~cm and shows a gradual increase over time. This radius should be regarded as a lower limit on the emitting radius.
If this component originates from pre-existing circumstellar dust present at the time of explosion, it must correspond to dust that survived sublimation by the SN radiation. 
The dust sublimation radius can be estimated by following \citet{fox2011};

\begin{equation}
r_{\rm evap}
=
\left(
\frac{L_{\rm peak}}
{16 \pi \sigma T_{\rm evap}^{4} \langle Q \rangle}
\right)^{1/2}.
\label{eq:revap}
\end{equation}

Here, $T_{\rm evap}$ is the dust sublimation temperature and
$\langle Q \rangle$ is the Planck-averaged absorption coefficient of the dust.
In this case, assuming a dust sublimation temperature of $\sim1900$~K for carbon dust and a peak SN luminosity of $\sim10^{43}\,\mathrm{erg\,s^{-1}}$, the dust sublimation radius is estimated to be $\mathrm{r_{evap}}\sim1.5\times10^{17}$~cm.


The dust sublimation radius, $r_{\rm evap}\sim1.5\times10^{17}$~cm, corresponds to a light-travel time of $\sim58$~d. Therefore, if the NIR excess were produced by an IR echo from pre-existing dust located near this radius, the emission could in principle begin at relatively early phases. Interpreting the onset of the NIR excess at $\sim300$~d as an IR echo would naively suggest a characteristic dust-shell radius of $R\sim ct\sim7.8\times10^{17}$~cm. However, a detached CSM dust distribution with $R\gg r_{\rm evap}$ is not itself implausible, since pre-existing dust may reside in a shell or cavity formed by previous mass loss. In addition, the onset of an echo from the near side of a CSM shell can occur earlier than the full light-crossing time of $2R/c$. Thus, these geometrical considerations alone do not rule out an IR-echo scenario. The main difficulty with this interpretation is that the onset of the NIR excess coincides with the accelerated optical fading and the red-wing attenuation of H$\alpha$, both of which are more naturally associated with newly formed dust. We therefore disfavor an IR-echo scenario from pre-existing dust as the dominant origin of the observed NIR excess.

\subsubsection{Dust formation} \label{sssec:dust formation}
We now consider a scenario in which the observed IR excess originates from newly formed dust. As summarized in Sections~\ref{sec:photometry} and \ref{sec:spectra}, SN~2024dy exhibits several key observational signatures:~(i) an accelerated decline in the optical luminosity (Figures~\ref{multibandlc} and \ref{fig:lbol_fraction}), (ii) a rise in the NIR luminosity accompanied by a clear NIR excess (Figure~\ref{bbfit}), and (iii) a suppression of the red wing in line profiles, most prominently in H$\alpha$ (Figures~\ref{fig:ha_norma} and \ref{fig:ha_reverse}). These features become particularly pronounced during Season~2, and all of them can be naturally explained by the formation of newly formed dust \citep{smith2012}.

The increase in the optical decline rate from Season~1 to Season~2 suggests the emergence of additional obscuration along the line of sight. This is consistent with attenuation by newly formed dust. In this scenario, the absorbed optical radiation is re-emitted in the IR, which naturally explains the rise in the NIR luminosity. The suppression of the red wing in the H$\alpha$ profile provides further evidence of dust formation, as similar evolution has been reported in other long-lived Type~IIn SNe \citep[e.g.,][]{chugai2018, smith2024, reynolds2025}.

Although \citet{reynolds2025} discusses the possibility that red-wing attenuation may result from occultation by an optically thick photosphere, this explanation appears unlikely for SN~2024dy in Season~2. If photospheric occultation were responsible, the attenuation would be expected to weaken as the photosphere recedes. However, the H$\alpha$ profiles in Figure~10 show no clear temporal decrease in the suppression, favoring dust formation as the dominant origin of the Season~2 asymmetry.

Furthermore, as shown in the lower panel of Figure~\ref{fig:hafit_result}, the higher H$\alpha$/H$\beta$ ratio during Season~2 is also consistent with wavelength-dependent attenuation expected from dust absorption. Although the Balmer decrement can be influenced by several other physical processes, such as high optical depth, collisional excitation, or deviations from Case~B, this trend provides additional support for the dust-formation scenario.

When these observational characteristics are considered together, they strongly support the presence of newly formed dust in SN~2024dy.

The suppression of the red wing of the late-time H$\alpha$ profile indicates selective attenuation of the receding side emission along the line of sight, which is a well-known observational signature associated with dust formation.
Previous studies have interpreted red-wing suppression differently depending on the location of dust formation.
For example, \citet{chugai2018} compared radiative transfer models with the H$\alpha$ profile of SN~2010jl and argued that dust confined solely to the CDS could not reproduce the observed asymmetry. In contrast, \citet{smith2024} found in SN~2015da that the broad component remains largely symmetric, while only the intermediate-width component becomes asymmetric; together with a flux deficit near zero velocity, this favored the CDS as the dust formation site. Consistent with this view, numerical calculations by \citet{dessart2025} demonstrate that CDS dust can produce similar line-profile asymmetries.

The H$\alpha$ profile of SN~2024dy exhibits characteristics intermediate between these cases. Although red wing suppression is pronounced, little or no flux deficit appears near zero velocity (Figure~\ref{fig:ha_reverse}), suggesting selective attenuation of the receding emission only. Such a morphology can be reproduced by dust located either within the SN ejecta or in the CDS, provided that the dust distribution is asymmetric or clumpy. Therefore, the line profile alone does not uniquely constrain the dust formation site. 

The spectral evolution suggests that ejecta dust is unlikely to dominate the observed NIR emission. An additional important constraint is that, at least until the middle of Season~2, the spectra do not show clear ejecta-origin features such as nebular lines (e.g., O~{\sc i}~$\lambda\lambda6300,6364$). This suggests that the photosphere remains embedded within the CSM and that the SN ejecta are not yet exposed. Under such conditions, even if dust were forming within the ejecta, its thermal emission would not be expected to dominate the observed NIR luminosity. It is therefore difficult to attribute the NIR excess detected during Season~2 solely to ejecta dust.

These considerations suggest that dust formation in the CDS is the more plausible scenario for SN~2024dy. SN~2024dy shows evidence that strong interaction persists up to late phases ($\sim$500 d), implying that the CDS formed in the interaction region provides a favorable environment for dust formation over an extended period. 

\subsection{Estimation of dust mass} \label{ssec:estimation dustmass}

\subsubsection{SED modeling} \label{sssec:dustmodel}
In some previously observed Type IIn SNe, IR emission attributed to dust has been reported (e.g., SN~2005ip;~\citealt{fox2009}, SN~2010jl;~\citealt{fransson2014}, SN~2014ab;~\citealt{moriya2020}, SN~2017hcc;~\citealt{moran2023}). In particular, SN~2010jl exhibited an IR excess caused by dust emission in its SED from an early phase. The NIR excess observed in SN~2024dy in Season~2 (see section~\ref{ssec:bbfit}) is also likely to originate from newly formed dust, similar to these cases.

Here, we modeled the cool component of the SED with dust emission calculated from dust emission models. The dust emission is described by
\begin{equation}
F_{\lambda,{\rm dust}}=\frac{M_{\rm dust}\,B_\lambda(T_{\rm dust})\kappa_\lambda(a)}{D^2}
\end{equation}
where $M_{\rm dust}$ is total dust mass, $B_\lambda(T_{\rm dust})$ is the BB emission at the dust temperature $T_{\rm dust}$, $\kappa_\lambda(a)$ is the mass absorption coefficient for the dust grains of radius $a$ as derived from Mie theory, and $D$ is the distance to the emission source. We investigated the temporal evolution of the dust temperature and mass that best reproduces the observed SED.
The models \citep[see][for details]{kawabata2007} assume optically thin dust emission, with two possible compositions: amorphous carbon ('BE' soot case from~\citet{rouleau1991}) and astronomical silicate~\citep{draine1985}. We assume a single grain size of $0.01~\mu\mathrm{m}$, and explore dust masses in the range $10^{-6}$--$10^{-3}\,M_{\odot}$. Dust emission is calculated for temperatures between $500$ and $1900~\mathrm{K}$ in steps of 10~K.

For each pair of $T_{\rm dust}$ and $M_{\rm dust}$, we constructed a BB+dust model by adding a hot BB component to the pre-computed dust-emission SED. The temperature and radius of the hot BB component were treated as free parameters and optimized by least-squares fitting. We then calculated $\chi^2$ for each BB+dust combination and adopted the model with the minimum $\chi^2$ as the best-fit solution.

\begin{figure}[htbp]
    \centering
    \includegraphics[width=0.49\textwidth]{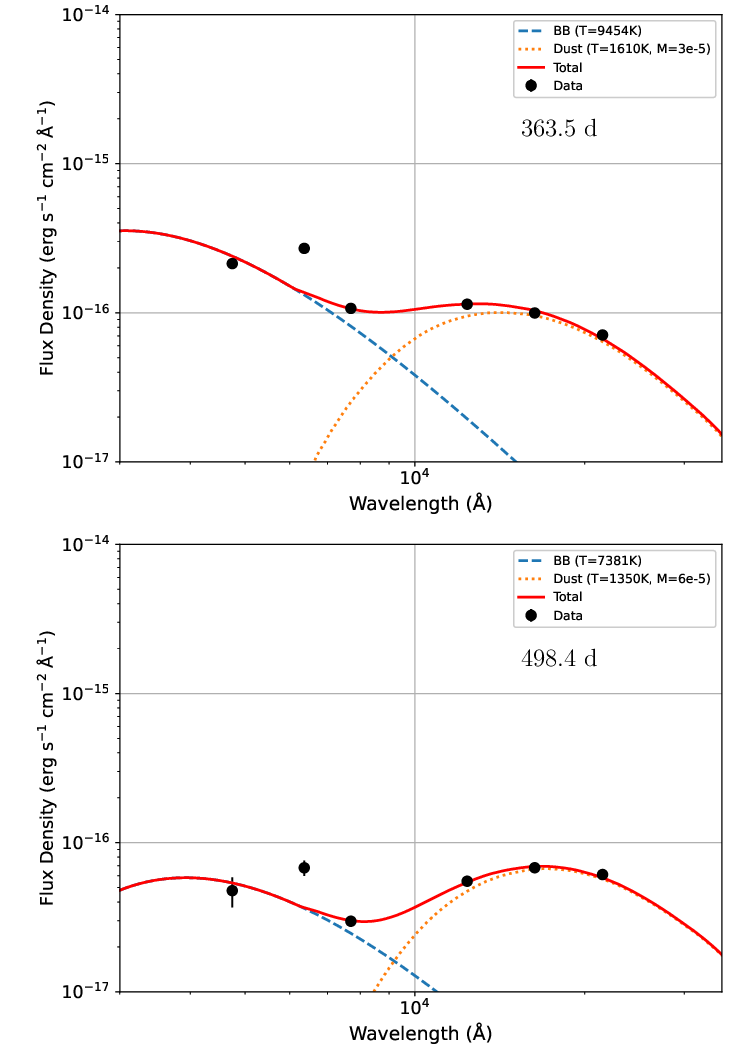}
    \caption{BB and dust component fits to the SED at representative epochs. These panels present selected epochs from Season~2, where the SEDs are modeled with a combination of a BB component and carbon dust emission (orange dashed lines). The red solid lines indicate the total model SED. The $r$-band is excluded from the fitting due to contamination from H$\alpha$ emission.}
    \label{dustfit}
\end{figure}

Figure~\ref{dustfit} presents the SEDs at representative epochs, including the final epoch of Season~1 and two epochs from Season~2. A single BB component reproduces the Season~1 SED well (as discussed in section~\ref{ssec:bbfit}). For the Season~2 epochs, the observed NIR excess is well explained by adding a dust emission component, modeled as carbon dust, to the BB emission. The $r$-band is excluded from the fitting because it is contaminated by H$\alpha$ emission. Through Season~2, the flux of the BB component (dominating the optical region) declines with time, while the NIR dust emission remains nearly constant, with its peak gradually shifting toward longer wavelengths.

Table~\ref{tab:dust_param} summarizes the temporal evolution of the dust temperature and mass inferred from dust emission models assuming carbon grains. At the beginning of Season~2, the dust temperature is estimated to be $\sim1800$~K, then the temperature decreases to $\sim1350$~K over the subsequent $\sim200$ days. These temperatures are well below the typical sublimation temperature of carbon dust ($\sim1900$~K; e.g., \citealt{dwek1985}). In contrast, fitting the data with silicate compositions yields temperatures that exceed the silicate sublimation temperature ($\sim1500$~K; e.g., \citealt{dwek1985}) at many epochs, rendering such models physically implausible. Therefore, we conclude that the dust responsible for the observed NIR excess is likely dominated by carbon grains rather than silicates.

\begin{table}[htbp]
\centering
\caption{Estimated carbon dust temperature ($T_{\rm dust}$) and dust mass ($M_{\rm dust}$) at each epoch}
\label{tab:dust_param}
\begin{tabular}{c c c c}
\hline
MJD & Epoch (d) & $T_{\rm dust}$ (K) & $M_{\rm dust}$ ($M_\odot$) \\
\hline
60594.8 & 285.6 & $1840^{+25}_{-14}$ & $(1.0^{+0.1}_{-0.2}) \times 10^{-5}$ \\
60617.8 & 308.6 & $1880^{+20}_{-15}$ & $(1.0^{+0.7}_{-0.1}) \times 10^{-5}$ \\
60625.7 & 316.5 & $1640^{+9}_{-17}$ & $(2.0^{+0.3}_{-0.1}) \times 10^{-5}$ \\
60626.8 & 317.6 & $1680^{+17}_{-9}$ & $(2.0^{+0.1}_{-0.1}) \times 10^{-5}$ \\
60635.8 & 326.6 & $1710^{+11}_{-13}$ & $(2.0^{+0.2}_{-0.1}) \times 10^{-5}$ \\
60672.7 & 363.5 & $1610^{+9}_{-11}$ & $(3.0^{+1.2}_{-0.1}) \times 10^{-5}$ \\
60693.8 & 384.6 & $1460^{+7}_{-10}$ & $(5.0^{+1.3}_{-0.5}) \times 10^{-5}$ \\
60695.8 & 386.6 & $1510^{+10}_{-4}$ & $(4.0^{+0.4}_{-0.2}) \times 10^{-5}$ \\
60732.7 & 423.5 & $1450^{+5}_{-3}$ & $(5.0^{+0.5}_{-0.4}) \times 10^{-5}$ \\
60753.6 & 444.4 & $1430^{+6}_{-4}$ & $(5.0^{+0.4}_{-0.5}) \times 10^{-5}$ \\
60754.7 & 445.5 & $1430^{+2}_{-10}$ & $(5.0^{+0.5}_{-0.4}) \times 10^{-5}$ \\
60795.5 & 486.3 & $1390^{+5}_{-2}$ & $(5.0^{+1.1}_{-0.2}) \times 10^{-5}$ \\
60807.6 & 498.4 & $1350^{+8}_{-2}$ & $(6.0^{+1.3}_{-0.3}) \times 10^{-5}$ \\
\hline
\end{tabular}
\tablecomments{
The uncertainties correspond to the projected ranges of the 1-sigma confidence region in the chi-squared values for varying $T_{\rm{dust}}$ and $M_{\rm{dust}}$.}
\end{table}

\subsubsection{Effects of Optical Depth on Dust Mass Estimates} \label{sssec:optical depth effects}
In Section~\ref{sssec:dustmodel}, we applied dust emission models to the NIR excess and estimated the temperature and mass of the dust. However, these estimates may not capture the total amount of dust that formed. In this section, we discuss why the inferred dust mass may be an underestimate of the total dust mass produced in SN~2024dy.

\begin{figure}[htbp]
    \centering
    \includegraphics[width=0.5\textwidth]{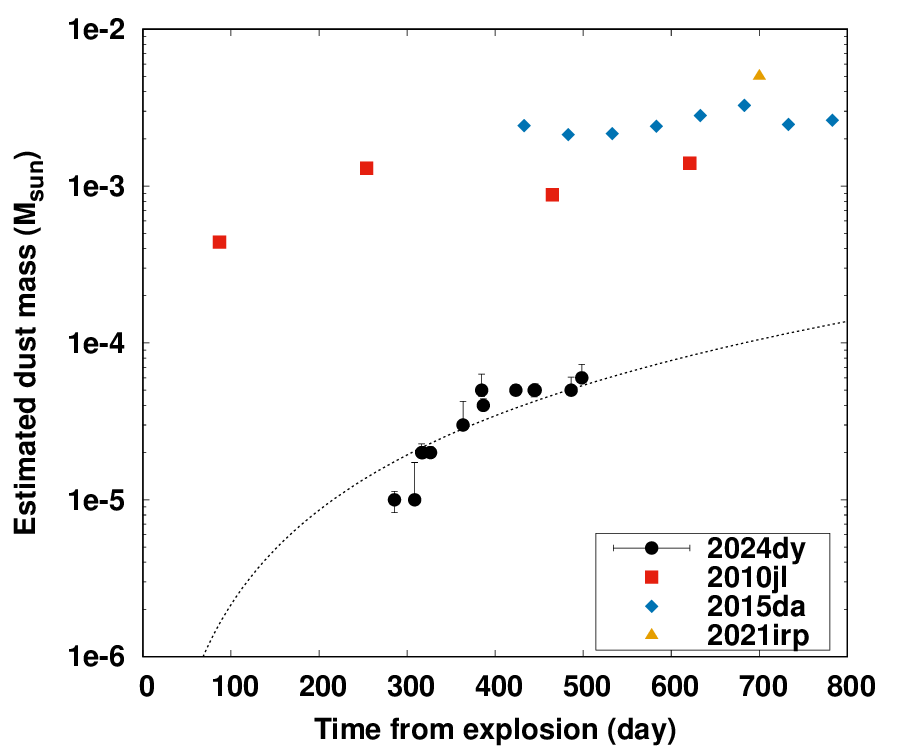}
    \caption{Temporal evolution of the dust mass inferred for SN~2024dy (black points), derived from the NIR excess assuming optically thin emission from hot dust (Section~\ref{sssec:dustmodel}). For comparison, the dust mass estimates for SN~2010jl~\citep{sarangi2018}, SN~2015da~\citep{tartaglia2020}, and SN~2021irp~\citep{reynolds2025}, all assuming carbon dust, are also shown. The black dotted line indicates an arbitrarily scaled $M_{\rm dust, obs} \propto t^{2}$ relation expected from the evolution of the effective emitting surface of optically thick dust, as discussed by~\citep{dwek2019}, and is shown for illustrative purposes only. The systematically lower dust mass inferred for SN~2024dy and its apparent consistency with the $t^2$ trend suggest that the derived dust mass may represent a lower limit rather than the total amount of dust formed.}
    \label{fig:dustmass_compa_all}
\end{figure}

Figure~\ref{fig:dustmass_compa_all} shows the time evolution of the dust mass we estimated for SN~2024dy in Section~\ref{sssec:dustmodel} (black points). Similarly, we show for comparison the time evolution of dust masses estimated under the assumption of carbon grains for several other Type~II SNe, including SN~2010jl \citep{sarangi2018}, SN~2015da \citep{tartaglia2020}, and SN~2021irp \citep{reynolds2025}. As is clear from this comparison, the dust mass inferred for SN~2024dy is noticeably smaller than the estimates for other objects with similar characteristics. In the following, we consider whether the dust mass in SN~2024dy could be underestimated, while noting that similar systematic effects may also affect the dust-mass estimates for the comparison objects.

\citet{dwek2019} pointed out that the apparent increase in dust mass with time can arise from the decreasing IR optical depth of expanding, optically thick ejecta, rather than from slow dust formation. In this picture, most of the dust may form rapidly at early phases, but only a fraction of its IR emission can escape. As the ejecta expand and the optical depth decreases, a larger fraction of formed dust becomes observable, producing an apparent increase in the inferred dust mass approximately following $M_{\rm dust, obs} \propto t^{2}$.
Thus, a small dust mass inferred at early epochs does not necessarily imply inefficient dust formation; rather, it may simply reflect the fact that the bulk of the dust in the inner, optically thick regions has not yet become observable. As shown in Figure~\ref{fig:dustmass_compa_all}, the relation $M_{\rm obs} \propto t^{2}$ (black dot line) appears to reproduce the time evolution of the dust mass estimated for SN~2024dy remarkably well. This suggests that the inferred value of $\sim10^{-5}\ M_{\odot}$ represents only the mass of the `apparently optically thin' dust, rather than the total amount of dust that was actually formed.

We also note that an additional possibility is the presence of cooler dust that would not be detectable in the NIR. In several dust-forming SNe such as SN~2010jl, the presence of an additional, cooler dust component ($\mathrm{T_{dust}}\lesssim1000$ K) has been suggested from MIR observations~\citep[e.g.,][]{fox2011}. However, our NIR data are not sensitive to such low-temperature dust emission, and we cannot place meaningful constraints on its presence or absence. A detailed investigation of this component is therefore beyond the scope of this paper.

These considerations imply that it is difficult to infer the amount of optical extinction directly from the dust masses estimated through the NIR emission. We therefore also estimated the dust mass responsible for the observed optical attenuation by evaluating the degree of optical fading, and compared this value with the dust masses inferred from the NIR excess. We estimated the optical depth $\tau$ required to reproduce two observational constraints: the suppression of the red wing in the H$\alpha$ profile and the accelerated decline of the optical pseudo-bolometric light curve.

From the flux deficit in the red wing of the last spectrum at $391$~d (Section~\ref{ssec:line evolution}), we derived $\tau \sim 0.9$. As an independent estimate, we compared the optical and optical+NIR pseudo-bolometric light curves in Figure~\ref{fig:lbol_fraction}. During Season~1 ($\sim100$--$280$~d), these two light curves show similar decline rates after a vertical scaling. At $391$~d, however, the optical luminosity is lower by a factor of $\sim2$--$3$ relative to the expected value extrapolated from the decline of Season~1. This corresponds to an additional fading of $\Delta m \sim 0.9$--$1.1$~mag, or $\tau \sim 0.8$--$1.0$. These two independent estimates are broadly consistent, suggesting an optical depth of order $\tau \sim 0.9$ during Season~2.

The optical depth is defined as
\begin{equation} 
\tau = \int \kappa \rho \, dr .
\label{equ4}
\end{equation}

Assuming a thin dust shell with mass $M_{\rm dust, abs}$ at radius $r$ and thickness $\Delta r$, the density is written as
\begin{equation}
\rho \sim \frac{M_{\rm dust, abs}}{4\pi r^{2} \Delta r}.
\label{equ5}
\end{equation}

Substituting this into Equation~\ref{equ4}, we obtain
\begin{equation}
\tau \sim \kappa \rho \Delta r
\sim \frac{\kappa M_{\rm dust, abs}}{4\pi r^{2}}.
\label{equ6}
\end{equation}

This expression can be rearranged to give the dust mass in terms of the optical depth. We rewrite Equation~\ref{equ6} as :
\begin{equation}
M_{\rm{dust, abs}}=\frac{4\pi{r_{BB}^2}}{\kappa}\tau
\label{equ7}
\end{equation}

Where $M_{\rm dust, abs}$ denotes the dust mass contributing to the absorption, and $r_{\rm BB}$ is the BB radius estimated from the SED fitting, which also provides a lower limit on the radial extent of the dust–emitting region. Therefore, the dust mass derived using $r_{\rm BB}$ should be regarded as a lower limit. The parameter $\kappa$ is the dust absorption coefficient, for which we adopt a representative value of $\kappa \sim 30000\ \mathrm{cm^{2}\,g^{-1}}$ at the wavelength of H$\alpha$~\citep{rouleau1991}. Using equation~\ref{equ7}, the dust mass required to account for the optical attenuation is estimated to be {$M_{\mathrm{dust,abs}} \geq 1.6 \times 10^{-4}\,{\rm M_{\odot}}$. This value is approximately an order of magnitude larger than the dust mass inferred from the SED fitting at the same epoch. We also note that this estimate focuses only on the dust responsible for the absorption, and thus represents a lower limit on the total dust mass that was actually formed. Additional dust may remain undetected owing to geometric asymmetry or other effects. 

These results imply that the NIR emission traces only a fraction of the dust actually formed at $\sim400$~d, indicating that a substantial amount of the newly formed dust remained missing in our view. This estimate is also consistent with the fact that the estimated dust mass from SED is well reproduced by the $M_{\rm dust, obs} \propto t^{2}$ relation proposed by \citet{dwek2019}, as discussed above. A similar argument has been made for many Type~IIP SNe \citep[e.g.,][]{Tinyanont2019}, indicating the need for more systematic NIR/MIR studies of Type~IIn SNe as well.

From these considerations, we conclude that the dust masses derived from the SED fitting represent only the `visible' dust component and trace only a fraction of the dust responsible for the optical attenuation. In addition, cooler dust may also be present but remains unconstrained by our NIR data. Therefore, the dust masses estimated in this work should be regarded as lower limits to the total amount of dust formed in SN~2024dy.

\section{Summary and conclusions} \label{sec:summary}
In this study, we present UV-Opt-NIR photometric and spectroscopic observations of the Type~IIn SN~2024dy, obtained over a period of approximately $500$~days. Below, we summarize the key results and discussions presented in this study.

(1) SN~2024dy is a long-lived Type~IIn monitored for nearly 500 days in the optical and NIR. Its peak magnitude is $M_r \simeq -19.2$~mag, consistent with the typical luminosity of this class. The rise time was $38$ d, placing it toward the longer end of the distribution of Type IIn SNe.

(2) After the solar conjunction, the Season~2 light curves exhibit a marked change in evolution. In the optical bands, the decline rate increases by a factor of about four compared to Season~1, whereas the NIR bands became nearly flat or showed signs of rebrightening.

(3) The Season~2 SEDs show a pronounced NIR excess, which is well reproduced by a two-component BB consisting of a hot component with $T\sim10^4$--$6\times10^3$~K and a cooler component with $T<3000$~K. The pseudo-bolometric light curve peaks at $1.6\times10^{43}\,\mathrm{erg\,s^{-1}}$, and the total radiated energy over $500$ days is $1.9\times10^{50}$~erg. During Season~2, the ratio $L_{\rm NIR}/L_{\rm opt}$ increases steadily with time.

(4) The early-time spectra are dominated by Balmer emission lines and a blue continuum, both characteristic of strong CSM interaction in Type~IIn SNe, and several He lines are also detected. After its peak, a blueshifted component appears in H$\alpha$, and by Season~2 the line evolves into an asymmetric profile with a suppressed red wing, indicative of dust obscuration. Throughout the entire monitoring period, no clear ejecta-origin features emerge, suggesting that strong CSM interaction persists to the late phases.

(5) The mass-loss rate inferred from the peak luminosity is $0.037\ M_\odot\,\mathrm{yr^{-1}}$. Although this value is somewhat lower than those of comparison objects, the comparable total radiated energy suggests that SN~2024dy was surrounded by a CSM mass similar to that of the comparison events, likely distributed over a more extended and lower density region.

(6) Assuming the NIR excess arises from optically thin dust emission, the inferred carbon dust temperature decreases from $1800$~K to $1300$~K, while the dust mass increases to $1$--$6\times10^{-5}\ M_\odot$. The temperatures rule out silicate grains, indicating that carbon dust is the dominant component.
    
(7) The properties of the observed dust emission cannot be explained by an IR echo and instead point to newly formed dust. The data do not allow us to uniquely identify the formation site, and distinguishing between the inner ejecta and the CDS is challenging with the present observations. The dust mass estimated from the SED modeling is significantly smaller than the mass estimated from the attenuation of H$\alpha$, implying that a substantial fraction of the newly formed dust is either buried in an optically thick region or exists as cooler dust that is not detectable at NIR.

Long-lived Type~IIn SNe such as SN~2024dy remain rare, and their physical nature is still not well understood. Future facilities, including Rubin Observatory~\citep{ivezi2019}, are expected to provide increasing opportunities to monitor such events over timescales exceeding $1000$~days~\citep[e.g.,][]{jacobson2025, shahbandeh2025}. Expanding the sample of these long-lived interacting SNe will be essential for developing a more complete understanding of the physics of CSM interaction.

\begin{acknowledgments}
We thank Nozomu Tominaga, Takashi Moriya, and Ryotaro Chiba for the discussion.
We are grateful to the graduate and undergraduate students of the Kagoshima University (KU) 1m group for their efforts in performing the optical and near-infrared observations. This work was supported by Grant-in-Aid for Scientific Research (C) 22K03676. The KU 1-m and Kanata telescopes are members of the Optical and Infrared Synergetic Telescopes for Education and Research (OISTER) program funded by the MEXT of Japan. D.K.S. acknowledges the support provided by DST-JSPS under grant No. DST/INT/JSPS/P363/2022. This work was supported by JSPS Bilateral Program Number JPJSBP 120227709. A.. acknowledges the support from the research project grant “Understanding the Dynamic Universe” funded by the Knut and Alice Wallenberg under Dnr KAW 2018.0067.
\end{acknowledgments}

\section*{Data Availability}
The photometric data and pseudo-bolometric luminosities used in this work are available as data behind the relevant figures. The data files are provided in CSV format, together with ReadMe files describing the file contents, column definitions, units, magnitude systems, telescopes, instruments, and data sources. The spectra will be uploaded to WISeREP.

\appendix

\section{Observation log} \label{aptab:sepclog}
In this appendix, we present a summary of the spectroscopic observations used in this study. Table~\ref{tab:speclog} provides the observational log, including the observation date, phase, instrument, wavelength coverage, exposure time, and resolution for each epoch.

\begin{table*}[ht]
    \centering
    \caption{Log of spectroscopic observations of SN~2024dy}
    \label{tab:speclog}
    \begin{tabular}{c c c c c c c}
        \hline \hline
        Date & MJD & Phase & Telescope + instrument & Spectral Range & Exposure Time & Resolution \\
         &  & (d) &  & (\AA) & (s) & \\
        \hline
        20240216 & 60356.9 & 47.7 & DOT + ADFOSC & 3500-7100 & 600  & 700 \\
        20240308 & 60377.8 & 68.6 & DOT + ADFOSC & 3500-7100 & 900  & 700 \\
        20240315 & 60384.8 & 75.6 & DOT + ADFOSC & 3500-7100 & 600  & 700 \\
        20240411 & 60411.6 & 102.4 & HCT + HFOSC & 3800-8000 & 2400 & $\sim 500$ \\
        20240424 & 60424.7 & 115.5 & HCT + HFOSC & 3800-8000 & 1800 & $\sim 500$ \\
        20240504 & 60434.7 & 125.5 & HCT + HFOSC & 3800-8000 & 2400 & $\sim 500$ \\
        20240514 & 60444.8 & 135.6 & HCT + HFOSC & 3800-8000 & 2700 & $\sim 500$ \\
        20240518 & 60448.7 & 139.5 & HCT + HFOSC & 3800-8000 & 2700 & $\sim 500$ \\
        20240604 & 60465.7 & 156.5 & HCT + HFOSC & 3800-8000 & 1050 & $\sim 500$ \\
        20240614 & 60475.6 & 166.4 & HCT + HFOSC & 3800-8000 & 1800 & $\sim 500$ \\
        20241114 & 60628.9 & 319.7 & HCT + HFOSC & 3800-8000 & 3600 & $\sim 500$ \\
        20241130 & 60644.9 & 335.7 & HCT + HFOSC & 3800-8000 & 2700 & $\sim 500$ \\
        20250105 & 60680.8 & 371.6 & HCT + HFOSC & 3800-8000 & 3600 & $\sim 500$ \\
        20250125 & 60700.8 & 391.6 & DOT + ADFOSC & 3500-7100 & 2700 & 700 \\
        \hline
    \end{tabular}
\end{table*}

\section{result of line fitting} \label{apfig:hafit}
Table~\ref{tab:ha_profile} summarizes the parameters obtained from the H$\alpha$ line-profile fitting shown in Figure~\ref{fig:hafit_result}.

\begin{table*}
\centering
\caption{FWHM measurements of the H$\alpha$ line profile of SN~2024dy}
\label{tab:ha_profile}
\begin{tabularx}{0.8\textwidth}{@{\extracolsep{\fill}} c |c c c c}
\hline
Epoch & Gaussian 1 & Gaussian 2 & Gaussian 3 & Lorentzian \\
& (km $\mathrm{s^{-1}}$) & (km $\mathrm{s^{-1}}$) & (km $\mathrm{s^{-1}}$) & (km $\mathrm{s^{-1}}$) \\
\hline
8.4   & $346.7 \pm 7.4$   & --                  & --                  & $1643.7 \pm 46.4$ \\
9.1   & $785.2 \pm 70.5$  & --                  & --                  & $1580.0 \pm 209.4$ \\
47.7  & $455.4 \pm 16.9$  & --                  & --                  & $1865.2 \pm 83.9$ \\
68.6  & $443.5 \pm 15.7$  & $8067.0 \pm 278.5$  & $1804.6 \pm 61.9$  & -- \\
75.6  & $490.9 \pm 28.6$  & $8076.5 \pm 315.4$  & $1825.3 \pm 88.2$  & -- \\
102.4 & $564.4 \pm 24.6$  & $7282.5 \pm 79.9$   & $1822.3 \pm 41.1$  & -- \\
115.5 & $443.1 \pm 22.7$  & $6647.5 \pm 87.4$   & $1778.5 \pm 38.8$  & -- \\
125.5 & $626.0 \pm 36.5$  & $7637.5 \pm 68.6$   & $1810.9 \pm 41.3$  & -- \\
135.6 & $585.8 \pm 50.6$  & $7598.9 \pm 83.9$   & $1752.3 \pm 45.9$  & -- \\
139.5 & $565.8 \pm 39.6$  & $7881.1 \pm 81.0$   & $1839.7 \pm 43.9$  & -- \\
156.5 & $379.5 \pm 55.4$  & $6818.9 \pm 80.0$   & $1658.8 \pm 45.2$  & -- \\
166.4 & $449.1 \pm 52.5$  & $7084.3 \pm 79.9$   & $1731.3 \pm 43.3$  & -- \\
319.7 & $719.5 \pm 82.7$  & $6097.7 \pm 80.1$   & $1514.8 \pm 67.2$  & -- \\
335.7 & $1083.0 \pm 94.2$ & $5975.0 \pm 46.9$   & $1578.0 \pm 166.7$ & -- \\
371.6 & $964.0 \pm 212.7$ & $5067.4 \pm 86.8$   & $1415.8 \pm 301.9$ & -- \\
391.6 & $779.5 \pm 114.3$ & $4934.4 \pm 127.6$  & $1554.2 \pm 75.0$  & -- \\
\hline
\end{tabularx}
\tablecomments{Each epoch is given in days since the estimated explosion date (MJD~$60309.2$). The first three spectra are well reproduced by a combination of a narrow Gaussian component and an intermediate Lorentzian component.}
\end{table*}

\bibliography{2024dy}{}

@ARTICLE{tonry2018,
       author = {{Tonry}, J.~L. and {Denneau}, L. and {Heinze}, A.~N. and {Stalder}, B. and {Smith}, K.~W. and {Smartt}, S.~J. and {Stubbs}, C.~W. and {Weiland}, H.~J. and {Rest}, A.},
        title = "{ATLAS: A High-cadence All-sky Survey System}",
      journal = {\pasp},
         year = 2018,
        month = jun,
       volume = {130},
       number = {988},
        pages = {064505},
          doi = {10.1088/1538-3873/aabadf},
archivePrefix = {arXiv},
       eprint = {1802.00879},
 primaryClass = {astro-ph.IM},
       adsurl = {https://ui.adsabs.harvard.edu/abs/2018PASP..130f4505T}
}

@ARTICLE{classification_wise2024,
       author = {{Wise}, J. and {Hinds}, K. and {Perley}, D. and {Bochenek}, O. and {Rich}, R.~M.},
        title = "{ZTF Transient Classification Report for 2024-01-08}",
      journal = {Transient Name Server Classification Report},
         year = 2024,
        month = jan,
       volume = {2024-82},
        pages = {1},
       adsurl = {https://ui.adsabs.harvard.edu/abs/2024TNSCR..82....1W}
}

@ARTICLE{siegel2014,
       author = {{Siegel}, Michael H. and {Porterfield}, Blair L. and {Linevsky}, Jacquelyn S. and {Bond}, Howard E. and {Holland}, Stephen T. and {Hoversten}, Erik A. and {Berrier}, Joshua L. and {Breeveld}, Alice A. and {Brown}, Peter J. and {Gronwall}, Caryl A.},
        title = "{The Swift UVOT Stars Survey. I. Methods and Test Clusters}",
      journal = {\aj},
         year = 2014,
        month = dec,
       volume = {148},
       number = {6},
          eid = {131},
        pages = {131},
          doi = {10.1088/0004-6256/148/6/131},
archivePrefix = {arXiv},
       eprint = {1408.4676},
 primaryClass = {astro-ph.SR},
       adsurl = {https://ui.adsabs.harvard.edu/abs/2014AJ....148..131S}
}

@ARTICLE{cardelli1989,
       author = {{Cardelli}, Jason A. and {Clayton}, Geoffrey C. and {Mathis}, John S.},
        title = "{The Relationship between Infrared, Optical, and Ultraviolet Extinction}",
      journal = {\apj},
         year = 1989,
        month = oct,
       volume = {345},
        pages = {245},
          doi = {10.1086/167900},
       adsurl = {https://ui.adsabs.harvard.edu/abs/1989ApJ...345..245C}
}

@ARTICLE{schlafly2011,
       author = {{Schlafly}, Edward F. and {Finkbeiner}, Douglas P.},
        title = "{Measuring Reddening with Sloan Digital Sky Survey Stellar Spectra and Recalibrating SFD}",
      journal = {\apj},
         year = 2011,
        month = aug,
       volume = {737},
       number = {2},
          eid = {103},
        pages = {103},
          doi = {10.1088/0004-637X/737/2/103},
archivePrefix = {arXiv},
       eprint = {1012.4804},
 primaryClass = {astro-ph.GA},
       adsurl = {https://ui.adsabs.harvard.edu/abs/2011ApJ...737..103S}
}

@ARTICLE{roming2005,
       author = {{Roming}, Peter W.~A. and {Kennedy}, Thomas E. and {Mason}, Keith O. and {Nousek}, John A. and {Ahr}, Lindy and {Bingham}, Richard E. and {Broos}, Patrick S. and {Carter}, Mary J. and {Hancock}, Barry K. and {Huckle}, Howard E. and {Hunsberger}, S.~D. and {Kawakami}, Hajime and {Killough}, Ronnie and {Koch}, T. Scott and {McLelland}, Michael K. and {Smith}, Kelly and {Smith}, Philip J. and {Soto}, Juan Carlos and {Boyd}, Patricia T. and {Breeveld}, Alice A. and {Holland}, Stephen T. and {Ivanushkina}, Mariya and {Pryzby}, Michael S. and {Still}, Martin D. and {Stock}, Joseph},
        title = "{The Swift Ultra-Violet/Optical Telescope}",
      journal = {\ssr},
         year = 2005,
        month = oct,
       volume = {120},
       number = {3-4},
        pages = {95-142},
          doi = {10.1007/s11214-005-5095-4},
archivePrefix = {arXiv},
       eprint = {astro-ph/0507413},
 primaryClass = {astro-ph},
       adsurl = {https://ui.adsabs.harvard.edu/abs/2005SSRv..120...95R}
}

@ARTICLE{stetson1987,
       author = {{Stetson}, Peter B.},
        title = "{DAOPHOT: A Computer Program for Crowded-Field Stellar Photometry}",
      journal = {\pasp},
         year = 1987,
        month = mar,
       volume = {99},
        pages = {191},
          doi = {10.1086/131977},
       adsurl = {https://ui.adsabs.harvard.edu/abs/1987PASP...99..191S}
}

@INPROCEEDINGS{nagayama2024,
       author = {{Nagayama}, Takahiro and {Nakaya}, Hidehiko},
        title = "{kSIRIUS: a simultaneous JHKs camera for the Kagoshima University 1m telescope using newly developed Japanese InGaAs array detectors}",
    booktitle = {Ground-based and Airborne Instrumentation for Astronomy X},
         year = 2024,
       editor = {{Bryant}, Julia J. and {Motohara}, Kentaro and {Vernet}, Jo{\"e}l. R.~D.},
       series = {Society of Photo-Optical Instrumentation Engineers (SPIE) Conference Series},
       volume = {13096},
        month = jul,
          eid = {130963I},
        pages = {130963I},
          doi = {10.1117/12.3016593},
       adsurl = {https://ui.adsabs.harvard.edu/abs/2024SPIE13096E..3IN}
}

@ARTICLE{chamber2016,
       author = {{Chambers}, K.~C. and {Magnier}, E.~A. and {Metcalfe}, N. and {Flewelling}, H.~A. and {Huber}, M.~E. and {Waters}, C.~Z. and {Denneau}, L. and {Draper}, P.~W. and {Farrow}, D. and {Finkbeiner}, D.~P. and {Holmberg}, C. and {Koppenhoefer}, J. and {Price}, P.~A. and {Rest}, A. and {Saglia}, R.~P. and {Schlafly}, E.~F. and {Smartt}, S.~J. and {Sweeney}, W. and {Wainscoat}, R.~J. and {Burgett}, W.~S. and {Chastel}, S. and {Grav}, T. and {Heasley}, J.~N. and {Hodapp}, K.~W. and {Jedicke}, R. and {Kaiser}, N. and {Kudritzki}, R. -P. and {Luppino}, G.~A. and {Lupton}, R.~H. and {Monet}, D.~G. and {Morgan}, J.~S. and {Onaka}, P.~M. and {Shiao}, B. and {Stubbs}, C.~W. and {Tonry}, J.~L. and {White}, R. and {Ba{\~n}ados}, E. and {Bell}, E.~F. and {Bender}, R. and {Bernard}, E.~J. and {Boegner}, M. and {Boffi}, F. and {Botticella}, M.~T. and {Calamida}, A. and {Casertano}, S. and {Chen}, W. -P. and {Chen}, X. and {Cole}, S. and {Deacon}, N. and {Frenk}, C. and {Fitzsimmons}, A. and {Gezari}, S. and {Gibbs}, V. and {Goessl}, C. and {Goggia}, T. and {Gourgue}, R. and {Goldman}, B. and {Grant}, P. and {Grebel}, E.~K. and {Hambly}, N.~C. and {Hasinger}, G. and {Heavens}, A.~F. and {Heckman}, T.~M. and {Henderson}, R. and {Henning}, T. and {Holman}, M. and {Hopp}, U. and {Ip}, W. -H. and {Isani}, S. and {Jackson}, M. and {Keyes}, C.~D. and {Koekemoer}, A.~M. and {Kotak}, R. and {Le}, D. and {Liska}, D. and {Long}, K.~S. and {Lucey}, J.~R. and {Liu}, M. and {Martin}, N.~F. and {Masci}, G. and {McLean}, B. and {Mindel}, E. and {Misra}, P. and {Morganson}, E. and {Murphy}, D.~N.~A. and {Obaika}, A. and {Narayan}, G. and {Nieto-Santisteban}, M.~A. and {Norberg}, P. and {Peacock}, J.~A. and {Pier}, E.~A. and {Postman}, M. and {Primak}, N. and {Rae}, C. and {Rai}, A. and {Riess}, A. and {Riffeser}, A. and {Rix}, H.~W. and {R{\"o}ser}, S. and {Russel}, R. and {Rutz}, L. and {Schilbach}, E. and {Schultz}, A.~S.~B. and {Scolnic}, D. and {Strolger}, L. and {Szalay}, A. and {Seitz}, S. and {Small}, E. and {Smith}, K.~W. and {Soderblom}, D.~R. and {Taylor}, P. and {Thomson}, R. and {Taylor}, A.~N. and {Thakar}, A.~R. and {Thiel}, J. and {Thilker}, D. and {Unger}, D. and {Urata}, Y. and {Valenti}, J. and {Wagner}, J. and {Walder}, T. and {Walter}, F. and {Watters}, S.~P. and {Werner}, S. and {Wood-Vasey}, W.~M. and {Wyse}, R.},
        title = "{The Pan-STARRS1 Surveys}",
      journal = {arXiv e-prints},
         year = 2016,
        month = dec,
          eid = {arXiv:1612.05560},
        pages = {arXiv:1612.05560},
          doi = {10.48550/arXiv.1612.05560},
archivePrefix = {arXiv},
       eprint = {1612.05560},
 primaryClass = {astro-ph.IM},
       adsurl = {https://ui.adsabs.harvard.edu/abs/2016arXiv161205560C}
}

@ARTICLE{skrutskie2006,
       author = {{Skrutskie}, M.~F. and {Cutri}, R.~M. and {Stiening}, R. and {Weinberg}, M.~D. and {Schneider}, S. and {Carpenter}, J.~M. and {Beichman}, C. and {Capps}, R. and {Chester}, T. and {Elias}, J. and {Huchra}, J. and {Liebert}, J. and {Lonsdale}, C. and {Monet}, D.~G. and {Price}, S. and {Seitzer}, P. and {Jarrett}, T. and {Kirkpatrick}, J.~D. and {Gizis}, J.~E. and {Howard}, E. and {Evans}, T. and {Fowler}, J. and {Fullmer}, L. and {Hurt}, R. and {Light}, R. and {Kopan}, E.~L. and {Marsh}, K.~A. and {McCallon}, H.~L. and {Tam}, R. and {Van Dyk}, S. and {Wheelock}, S.},
        title = "{The Two Micron All Sky Survey (2MASS)}",
      journal = {\aj},
         year = 2006,
        month = feb,
       volume = {131},
       number = {2},
        pages = {1163-1183},
          doi = {10.1086/498708},
       adsurl = {https://ui.adsabs.harvard.edu/abs/2006AJ....131.1163S}
}

@INPROCEEDINGS{akitaya2014,
       author = {{Akitaya}, Hiroshi and {Moritani}, Yuki and {Ui}, Takahiro and {Urano}, Takeshi and {Ohashi}, Yuma and {Kawabata}, Koji S. and {Nakashima}, Asami and {Sasada}, Mahito and {Sakimoto}, Kiyoshi and {Harao}, Tatsuya and {Miyamoto}, Hisashi and {Matsui}, Rieko and {Itoh}, Ryosuke and {Takaki}, Katsutoshi and {Ueno}, Issei and {Ohsugi}, Takashi and {Nakaya}, Hidehiko and {Yamashita}, Takuya and {Yoshida}, Michitoshi},
        title = "{HONIR: an optical and near-infrared simultaneous imager, spectrograph, and polarimeter for the 1.5-m Kanata telescope}",
    booktitle = {Ground-based and Airborne Instrumentation for Astronomy V},
         year = 2014,
       editor = {{Ramsay}, Suzanne K. and {McLean}, Ian S. and {Takami}, Hideki},
       series = {Society of Photo-Optical Instrumentation Engineers (SPIE) Conference Series},
       volume = {9147},
        month = aug,
          eid = {91474O},
        pages = {91474O},
          doi = {10.1117/12.2054577},
       adsurl = {https://ui.adsabs.harvard.edu/abs/2014SPIE.9147E..4OA}
}

@ARTICLE{Bell2019ZTF,
       author = {{Bellm}, Eric C. and {Kulkarni}, Shrinivas R. and {Graham}, Matthew J. and {Dekany}, Richard and {Smith}, Roger M. and {Riddle}, Reed and {Masci}, Frank J. and {Helou}, George and {Prince}, Thomas A. and {Adams}, Scott M. and {Barbarino}, C. and {Barlow}, Tom and {Bauer}, James and {Beck}, Ron and {Belicki}, Justin and {Biswas}, Rahul and {Blagorodnova}, Nadejda and {Bodewits}, Dennis and {Bolin}, Bryce and {Brinnel}, Valery and {Brooke}, Tim and {Bue}, Brian and {Bulla}, Mattia and {Burruss}, Rick and {Cenko}, S. Bradley and {Chang}, Chan-Kao and {Connolly}, Andrew and {Coughlin}, Michael and {Cromer}, John and {Cunningham}, Virginia and {De}, Kishalay and {Delacroix}, Alex and {Desai}, Vandana and {Duev}, Dmitry A. and {Eadie}, Gwendolyn and {Farnham}, Tony L. and {Feeney}, Michael and {Feindt}, Ulrich and {Flynn}, David and {Franckowiak}, Anna and {Frederick}, S. and {Fremling}, C. and {Gal-Yam}, Avishay and {Gezari}, Suvi and {Giomi}, Matteo and {Goldstein}, Daniel A. and {Golkhou}, V. Zach and {Goobar}, Ariel and {Groom}, Steven and {Hacopians}, Eugean and {Hale}, David and {Henning}, John and {Ho}, Anna Y.~Q. and {Hover}, David and {Howell}, Justin and {Hung}, Tiara and {Huppenkothen}, Daniela and {Imel}, David and {Ip}, Wing-Huen and {Ivezi{\'c}}, {\v{Z}}eljko and {Jackson}, Edward and {Jones}, Lynne and {Juric}, Mario and {Kasliwal}, Mansi M. and {Kaspi}, S. and {Kaye}, Stephen and {Kelley}, Michael S.~P. and {Kowalski}, Marek and {Kramer}, Emily and {Kupfer}, Thomas and {Landry}, Walter and {Laher}, Russ R. and {Lee}, Chien-De and {Lin}, Hsing Wen and {Lin}, Zhong-Yi and {Lunnan}, Ragnhild and {Giomi}, Matteo and {Mahabal}, Ashish and {Mao}, Peter and {Miller}, Adam A. and {Monkewitz}, Serge and {Murphy}, Patrick and {Ngeow}, Chow-Choong and {Nordin}, Jakob and {Nugent}, Peter and {Ofek}, Eran and {Patterson}, Maria T. and {Penprase}, Bryan and {Porter}, Michael and {Rauch}, Ludwig and {Rebbapragada}, Umaa and {Reiley}, Dan and {Rigault}, Mickael and {Rodriguez}, Hector and {van Roestel}, Jan and {Rusholme}, Ben and {van Santen}, Jakob and {Schulze}, S. and {Shupe}, David L. and {Singer}, Leo P. and {Soumagnac}, Maayane T. and {Stein}, Robert and {Surace}, Jason and {Sollerman}, Jesper and {Szkody}, Paula and {Taddia}, F. and {Terek}, Scott and {Van Sistine}, Angela and {van Velzen}, Sjoert and {Vestrand}, W. Thomas and {Walters}, Richard and {Ward}, Charlotte and {Ye}, Quan-Zhi and {Yu}, Po-Chieh and {Yan}, Lin and {Zolkower}, Jeffry},
        title = "{The Zwicky Transient Facility: System Overview, Performance, and First Results}",
      journal = {\pasp},
         year = 2019,
        month = jan,
       volume = {131},
       number = {995},
        pages = {018002},
          doi = {10.1088/1538-3873/aaecbe},
archivePrefix = {arXiv},
       eprint = {1902.01932},
 primaryClass = {astro-ph.IM},
       adsurl = {https://ui.adsabs.harvard.edu/abs/2019PASP..131a8002B}
}

@ARTICLE{foster2021,
       author = {{F{\"o}rster}, F. and {Cabrera-Vives}, G. and {Castillo-Navarrete}, E. and {Est{\'e}vez}, P.~A. and {S{\'a}nchez-S{\'a}ez}, P. and {Arredondo}, J. and {Bauer}, F.~E. and {Carrasco-Davis}, R. and {Catelan}, M. and {Elorrieta}, F. and {Eyheramendy}, S. and {Huijse}, P. and {Pignata}, G. and {Reyes}, E. and {Reyes}, I. and {Rodr{\'\i}guez-Mancini}, D. and {Ruz-Mieres}, D. and {Valenzuela}, C. and {{\'A}lvarez-Maldonado}, I. and {Astorga}, N. and {Borissova}, J. and {Clocchiatti}, A. and {De Cicco}, D. and {Donoso-Oliva}, C. and {Hern{\'a}ndez-Garc{\'\i}a}, L. and {Graham}, M.~J. and {Jord{\'a}n}, A. and {Kurtev}, R. and {Mahabal}, A. and {Maureira}, J.~C. and {Mu{\~n}oz-Arancibia}, A. and {Molina-Ferreiro}, R. and {Moya}, A. and {Palma}, W. and {P{\'e}rez-Carrasco}, M. and {Protopapas}, P. and {Romero}, M. and {Sabatini-Gacitua}, L. and {S{\'a}nchez}, A. and {San Mart{\'\i}n}, J. and {Sep{\'u}lveda-Cobo}, C. and {Vera}, E. and {Vergara}, J.~R.},
        title = "{The Automatic Learning for the Rapid Classification of Events (ALeRCE) Alert Broker}",
      journal = {\aj},
         year = 2021,
        month = may,
       volume = {161},
       number = {5},
          eid = {242},
        pages = {242},
          doi = {10.3847/1538-3881/abe9bc},
archivePrefix = {arXiv},
       eprint = {2008.03303},
 primaryClass = {astro-ph.IM},
       adsurl = {https://ui.adsabs.harvard.edu/abs/2021AJ....161..242F}
}

@ARTICLE{smith2020atlas,
       author = {{Smith}, K.~W. and {Smartt}, S.~J. and {Young}, D.~R. and {Tonry}, J.~L. and {Denneau}, L. and {Flewelling}, H. and {Heinze}, A.~N. and {Weiland}, H.~J. and {Stalder}, B. and {Rest}, A. and {Stubbs}, C.~W. and {Anderson}, J.~P. and {Chen}, T. -W. and {Clark}, P. and {Do}, A. and {F{\"o}rster}, F. and {Fulton}, M. and {Gillanders}, J. and {McBrien}, O.~R. and {O'Neill}, D. and {Srivastav}, S. and {Wright}, D.~E.},
        title = "{Design and Operation of the ATLAS Transient Science Server}",
      journal = {\pasp},
         year = 2020,
        month = aug,
       volume = {132},
       number = {1014},
          eid = {085002},
        pages = {085002},
          doi = {10.1088/1538-3873/ab936e},
archivePrefix = {arXiv},
       eprint = {2003.09052},
 primaryClass = {astro-ph.IM},
       adsurl = {https://ui.adsabs.harvard.edu/abs/2020PASP..132h5002S}
}

@ARTICLE{heinze2018atlas,
       author = {{Heinze}, A.~N. and {Tonry}, J.~L. and {Denneau}, L. and {Flewelling}, H. and {Stalder}, B. and {Rest}, A. and {Smith}, K.~W. and {Smartt}, S.~J. and {Weiland}, H.},
        title = "{A First Catalog of Variable Stars Measured by the Asteroid Terrestrial-impact Last Alert System (ATLAS)}",
      journal = {\aj},
         year = 2018,
        month = nov,
       volume = {156},
       number = {5},
          eid = {241},
        pages = {241},
          doi = {10.3847/1538-3881/aae47f},
archivePrefix = {arXiv},
       eprint = {1804.02132},
 primaryClass = {astro-ph.SR},
       adsurl = {https://ui.adsabs.harvard.edu/abs/2018AJ....156..241H}
}

@ARTICLE{moran2023,
       author = {{Moran}, S. and {Fraser}, M. and {Kotak}, R. and {Pastorello}, A. and {Benetti}, S. and {Brennan}, S.~J. and {Guti{\'e}rrez}, C.~P. and {Kankare}, E. and {Kuncarayakti}, H. and {Mattila}, S. and {Reynolds}, T.~M. and {Anderson}, J.~P. and {Brown}, P.~J. and {Campana}, S. and {Chambers}, K.~C. and {Chen}, T. -W. and {Della Valle}, M. and {Dennefeld}, M. and {Elias-Rosa}, N. and {Galbany}, L. and {Galindo-Guil}, F.~J. and {Gromadzki}, M. and {Hiramatsu}, D. and {Inserra}, C. and {Leloudas}, G. and {M{\"u}ller-Bravo}, T.~E. and {Nicholl}, M. and {Reguitti}, A. and {Shahbandeh}, M. and {Smartt}, S.~J. and {Tartaglia}, L. and {Young}, D.~R.},
        title = "{A long life of excess: The interacting transient SN 2017hcc}",
      journal = {\aap},
         year = 2023,
        month = jan,
       volume = {669},
          eid = {A51},
        pages = {A51},
          doi = {10.1051/0004-6361/202244565},
archivePrefix = {arXiv},
       eprint = {2210.14076},
 primaryClass = {astro-ph.HE},
       adsurl = {https://ui.adsabs.harvard.edu/abs/2023A&A...669A..51M}
}

@ARTICLE{fransson2014,
       author = {{Fransson}, Claes and {Ergon}, Mattias and {Challis}, Peter J. and {Chevalier}, Roger A. and {France}, Kevin and {Kirshner}, Robert P. and {Marion}, G.~H. and {Milisavljevic}, Dan and {Smith}, Nathan and {Bufano}, Filomena and {Friedman}, Andrew S. and {Kangas}, Tuomas and {Larsson}, Josefin and {Mattila}, Seppo and {Benetti}, Stefano and {Chornock}, Ryan and {Czekala}, Ian and {Soderberg}, Alicia and {Sollerman}, Jesper},
        title = "{High-density Circumstellar Interaction in the Luminous Type IIn SN 2010jl: The First 1100 Days}",
      journal = {\apj},
         year = 2014,
        month = dec,
       volume = {797},
       number = {2},
          eid = {118},
        pages = {118},
          doi = {10.1088/0004-637X/797/2/118},
archivePrefix = {arXiv},
       eprint = {1312.6617},
 primaryClass = {astro-ph.HE},
       adsurl = {https://ui.adsabs.harvard.edu/abs/2014ApJ...797..118F}
}

@ARTICLE{fox2009,
       author = {{Fox}, Ori and {Skrutskie}, Michael F. and {Chevalier}, Roger A. and {Kanneganti}, Srikrishna and {Park}, Chan and {Wilson}, John and {Nelson}, Matthew and {Amirhadji}, Jason and {Crump}, Danielle and {Hoeft}, Alexi and {Provence}, Sydney and {Sargeant}, Benjamin and {Sop}, Joel and {Tea}, Matthew and {Thomas}, Steven and {Woolard}, Kyle},
        title = "{Near-Infrared Photometry of the Type IIn SN 2005ip: The Case for Dust Condensation}",
      journal = {\apj},
         year = 2009,
        month = jan,
       volume = {691},
       number = {1},
        pages = {650-660},
          doi = {10.1088/0004-637X/691/1/650},
archivePrefix = {arXiv},
       eprint = {0807.3555},
 primaryClass = {astro-ph},
       adsurl = {https://ui.adsabs.harvard.edu/abs/2009ApJ...691..650F}
}

@ARTICLE{ofek2014,
       author = {{Ofek}, Eran O. and {Arcavi}, Iair and {Tal}, David and {Sullivan}, Mark and {Gal-Yam}, Avishay and {Kulkarni}, Shrinivas R. and {Nugent}, Peter E. and {Ben-Ami}, Sagi and {Bersier}, David and {Cao}, Yi and {Cenko}, S. Bradley and {De Cia}, Annalisa and {Filippenko}, Alexei V. and {Fransson}, Claes and {Kasliwal}, Mansi M. and {Laher}, Russ and {Surace}, Jason and {Quimby}, Robert and {Yaron}, Ofer},
        title = "{Interaction-powered Supernovae: Rise-time versus Peak-luminosity Correlation and the Shock-breakout Velocity}",
      journal = {\apj},
         year = 2014,
        month = jun,
       volume = {788},
       number = {2},
          eid = {154},
        pages = {154},
          doi = {10.1088/0004-637X/788/2/154},
archivePrefix = {arXiv},
       eprint = {1404.4085},
 primaryClass = {astro-ph.HE},
       adsurl = {https://ui.adsabs.harvard.edu/abs/2014ApJ...788..154O}
}

@ARTICLE{nyholm2020,
       author = {{Nyholm}, A. and {Sollerman}, J. and {Tartaglia}, L. and {Taddia}, F. and {Fremling}, C. and {Blagorodnova}, N. and {Filippenko}, A.~V. and {Gal-Yam}, A. and {Howell}, D.~A. and {Karamehmetoglu}, E. and {Kulkarni}, S.~R. and {Laher}, R. and {Leloudas}, G. and {Masci}, F. and {Kasliwal}, M.~M. and {Mor{\r{a}}}, K. and {Moriya}, T.~J. and {Ofek}, E.~O. and {Papadogiannakis}, S. and {Quimby}, R. and {Rebbapragada}, U. and {Schulze}, S.},
        title = "{Type IIn supernova light-curve properties measured from an untargeted survey sample}",
      journal = {\aap},
         year = 2020,
        month = may,
       volume = {637},
          eid = {A73},
        pages = {A73},
          doi = {10.1051/0004-6361/201936097},
archivePrefix = {arXiv},
       eprint = {1906.05812},
 primaryClass = {astro-ph.SR},
       adsurl = {https://ui.adsabs.harvard.edu/abs/2020A&A...637A..73N}
}

@ARTICLE{hiramatsu2024,
       author = {{Hiramatsu}, Daichi and {Berger}, Edo and {Gomez}, Sebastian and {Blanchard}, Peter K. and {Kumar}, Harsh and {Athukoralalage}, Wasundara},
        title = "{Type IIn Supernovae. I. Uniform Light Curve Characterization and a Bimodality in the Radiated Energy Distribution}",
      journal = {arXiv e-prints},
         year = 2024,
        month = nov,
          eid = {arXiv:2411.07287},
        pages = {arXiv:2411.07287},
          doi = {10.48550/arXiv.2411.07287},
archivePrefix = {arXiv},
       eprint = {2411.07287},
 primaryClass = {astro-ph.HE},
       adsurl = {https://ui.adsabs.harvard.edu/abs/2024arXiv241107287H}
}

@ARTICLE{chugai1994,
       author = {{Chugai}, N.~N. and {Danziger}, I.~J.},
        title = "{SN 1988Z: low-mass ejecta colliding with the clumpy wind?}",
      journal = {\mnras},
         year = 1994,
        month = may,
       volume = {268},
        pages = {173-180},
          doi = {10.1093/mnras/268.1.173},
       adsurl = {https://ui.adsabs.harvard.edu/abs/1994MNRAS.268..173C}
}

@ARTICLE{reynolds2025,
       author = {{Reynolds}, T.~M. and {Nagao}, T. and {Gottumukkala}, R. and {Guti{\'e}rrez}, C.~P. and {Kangas}, T. and {Kravtsov}, T. and {Kuncarayakti}, H. and {Maeda}, K. and {Elias-Rosa}, N. and {Fraser}, M. and {Kotak}, R. and {Mattila}, S. and {Pastorello}, A. and {Pessi}, P.~J. and {Cai}, Y. -Z. and {Fynbo}, J.~P.~U. and {Kawabata}, M. and {Lundqvist}, P. and {Matilainen}, K. and {Moran}, S. and {Reguitti}, A. and {Taguchi}, K. and {Yamanaka}, M.},
        title = "{The bright long-lived Type II SN 2021irp powered by aspherical circumstellar material interaction (I): Revealing the energy source with photometry and spectroscopy}",
      journal = {arXiv e-prints},
         year = 2025,
        month = jan,
          eid = {arXiv:2501.13619},
        pages = {arXiv:2501.13619},
          doi = {10.48550/arXiv.2501.13619},
archivePrefix = {arXiv},
       eprint = {2501.13619},
 primaryClass = {astro-ph.HE},
       adsurl = {https://ui.adsabs.harvard.edu/abs/2025arXiv250113619R}
}

@ARTICLE{tartaglia2020,
       author = {{Tartaglia}, L. and {Pastorello}, A. and {Sollerman}, J. and {Fransson}, C. and {Mattila}, S. and {Fraser}, M. and {Taddia}, F. and {Tomasella}, L. and {Turatto}, M. and {Morales-Garoffolo}, A. and {Elias-Rosa}, N. and {Lundqvist}, P. and {Harmanen}, J. and {Reynolds}, T. and {Cappellaro}, E. and {Barbarino}, C. and {Nyholm}, A. and {Kool}, E. and {Ofek}, E. and {Gao}, X. and {Jin}, Z. and {Tan}, H. and {Sand}, D.~J. and {Ciabattari}, F. and {Wang}, X. and {Zhang}, J. and {Huang}, F. and {Li}, W. and {Mo}, J. and {Rui}, L. and {Xiang}, D. and {Zhang}, T. and {Hosseinzadeh}, G. and {Howell}, D.~A. and {McCully}, C. and {Valenti}, S. and {Benetti}, S. and {Callis}, E. and {Carracedo}, A.~S. and {Fremling}, C. and {Kangas}, T. and {Rubin}, A. and {Somero}, A. and {Terreran}, G.},
        title = "{The long-lived Type IIn SN 2015da: Infrared echoes and strong interaction within an extended massive shell}",
      journal = {\aap},
         year = 2020,
        month = mar,
       volume = {635},
          eid = {A39},
        pages = {A39},
          doi = {10.1051/0004-6361/201936553},
archivePrefix = {arXiv},
       eprint = {1908.08580},
 primaryClass = {astro-ph.HE},
       adsurl = {https://ui.adsabs.harvard.edu/abs/2020A&A...635A..39T}
}

@ARTICLE{brennan2024,
       author = {{Brennan}, S.~J. and {Schulze}, S. and {Lunnan}, R. and {Sollerman}, J. and {Yan}, L. and {Fransson}, C. and {Irani}, I. and {Melinder}, J. and {Chen}, T.-W. and {De}, K. and {Fremling}, C. and {Kim}, Y.-L. and {Perley}, D. and {Pessi}, P.~J. and {Drake}, A.~J. and {Graham}, M.~J. and {Laher}, R.~R. and {Masci}, F.~J. and {Purdum}, J. and {Rodriguez}, H.},
        title = "{SN 2021adxl: A luminous nearby interacting supernova in an extremely low-metallicity environment}",
      journal = {\aap},
         year = 2024,
        month = oct,
       volume = {690},
          eid = {A259},
        pages = {A259},
          doi = {10.1051/0004-6361/202349036},
archivePrefix = {arXiv},
       eprint = {2312.13280},
 primaryClass = {astro-ph.HE},
       adsurl = {https://ui.adsabs.harvard.edu/abs/2024A&A...690A.259B}
}

@ARTICLE{stoll2011,
       author = {{Stoll}, R. and {Prieto}, J.~L. and {Stanek}, K.~Z. and {Pogge}, R.~W. and {Szczygie{\l}}, D.~M. and {Pojma{\'n}ski}, G. and {Antognini}, J. and {Yan}, H.},
        title = "{SN 2010jl in UGC 5189: Yet Another Luminous Type IIn Supernova in a Metal-poor Galaxy}",
      journal = {\apj},
         year = 2011,
        month = mar,
       volume = {730},
       number = {1},
          eid = {34},
        pages = {34},
          doi = {10.1088/0004-637X/730/1/34},
archivePrefix = {arXiv},
       eprint = {1012.3461},
 primaryClass = {astro-ph.CO},
       adsurl = {https://ui.adsabs.harvard.edu/abs/2011ApJ...730...34S}
}

@ARTICLE{nagao2025,
       author = {{Nagao}, T. and {Reynolds}, T.~M. and {Kuncarayakti}, H. and {Cartier}, R. and {Mattila}, S. and {Maeda}, K. and {Sollerman}, J. and {Pessi}, P.~J. and {Anderson}, J.~P. and {Inserra}, C. and {Chen}, T.-W. and {Ferrari}, L. and {Fraser}, M. and {Young}, D.~R. and {Gromadzki}, M. and {Guti{\'e}rrez}, C.~P. and {Lundqvist}, P. and {Pignata}, G. and {M{\"u}ller-Bravo}, T.~E. and {Ragosta}, F. and {Reguitti}, A. and {Moran}, S. and {Gonz{\'a}lez-Ba{\~n}uelos}, M. and {Kopsacheili}, M. and {Petrushevska}, T.},
        title = "{Observational diversity of bright long-lived Type II supernovae}",
      journal = {\aap},
         year = 2025,
        month = jul,
       volume = {699},
          eid = {A283},
        pages = {A283},
          doi = {10.1051/0004-6361/202554988},
archivePrefix = {arXiv},
       eprint = {2504.01427},
 primaryClass = {astro-ph.HE},
       adsurl = {https://ui.adsabs.harvard.edu/abs/2025A&A...699A.283N}
}

@ARTICLE{yaron2017,
       author = {{Yaron}, O. and {Perley}, D.~A. and {Gal-Yam}, A. and {Groh}, J.~H. and {Horesh}, A. and {Ofek}, E.~O. and {Kulkarni}, S.~R. and {Sollerman}, J. and {Fransson}, C. and {Rubin}, A. and {Szabo}, P. and {Sapir}, N. and {Taddia}, F. and {Cenko}, S.~B. and {Valenti}, S. and {Arcavi}, I. and {Howell}, D.~A. and {Kasliwal}, M.~M. and {Vreeswijk}, P.~M. and {Khazov}, D. and {Fox}, O.~D. and {Cao}, Y. and {Gnat}, O. and {Kelly}, P.~L. and {Nugent}, P.~E. and {Filippenko}, A.~V. and {Laher}, R.~R. and {Wozniak}, P.~R. and {Lee}, W.~H. and {Rebbapragada}, U.~D. and {Maguire}, K. and {Sullivan}, M. and {Soumagnac}, M.~T.},
        title = "{Confined dense circumstellar material surrounding a regular type II supernova}",
      journal = {Nature Physics},
         year = 2017,
        month = feb,
       volume = {13},
       number = {5},
        pages = {510-517},
          doi = {10.1038/nphys4025},
archivePrefix = {arXiv},
       eprint = {1701.02596},
 primaryClass = {astro-ph.HE},
       adsurl = {https://ui.adsabs.harvard.edu/abs/2017NatPh..13..510Y}
}

@ARTICLE{khazov2016,
       author = {{Khazov}, D. and {Yaron}, O. and {Gal-Yam}, A. and {Manulis}, I. and {Rubin}, A. and {Kulkarni}, S.~R. and {Arcavi}, I. and {Kasliwal}, M.~M. and {Ofek}, E.~O. and {Cao}, Y. and {Perley}, D. and {Sollerman}, J. and {Horesh}, A. and {Sullivan}, M. and {Filippenko}, A.~V. and {Nugent}, P.~E. and {Howell}, D.~A. and {Cenko}, S.~B. and {Silverman}, J.~M. and {Ebeling}, H. and {Taddia}, F. and {Johansson}, J. and {Laher}, R.~R. and {Surace}, J. and {Rebbapragada}, U.~D. and {Wozniak}, P.~R. and {Matheson}, T.},
        title = "{Flash Spectroscopy: Emission Lines from the Ionized Circumstellar Material around <10-day-old Type II Supernovae}",
      journal = {\apj},
         year = 2016,
        month = feb,
       volume = {818},
       number = {1},
          eid = {3},
        pages = {3},
          doi = {10.3847/0004-637X/818/1/3},
archivePrefix = {arXiv},
       eprint = {1512.00846},
 primaryClass = {astro-ph.HE},
       adsurl = {https://ui.adsabs.harvard.edu/abs/2016ApJ...818....3K}
}

@ARTICLE{salmaso2025,
       author = {{Salmaso}, I. and {Cappellaro}, E. and {Tartaglia}, L. and {Anderson}, J.~P. and {Benetti}, S. and {Bronikowski}, M. and {Cai}, Y.-Z. and {Charalampopoulos}, P. and {Chen}, T.-W. and {Concepcion}, E. and {Elias-Rosa}, N. and {Galbany}, L. and {Gromadzki}, M. and {Guti{\'e}rrez}, C.~P. and {Kankare}, E. and {Lundqvist}, P. and {Matilainen}, K. and {Mazzali}, P.~A. and {Moran}, S. and {M{\"u}ller-Bravo}, T.~E. and {Nicholl}, M. and {Pastorello}, A. and {Pessi}, P.~J. and {Pessi}, T. and {Petrushevska}, T. and {Pignata}, G. and {Reguitti}, A. and {Sollerman}, J. and {Srivastav}, S. and {Stritzinger}, M. and {Tomasella}, L. and {Valerin}, G.},
        title = "{The diversity of strongly interacting Type IIn supernovae}",
      journal = {\aap},
         year = 2025,
        month = mar,
       volume = {695},
          eid = {A29},
        pages = {A29},
          doi = {10.1051/0004-6361/202451764},
archivePrefix = {arXiv},
       eprint = {2410.06111},
 primaryClass = {astro-ph.HE},
       adsurl = {https://ui.adsabs.harvard.edu/abs/2025A&A...695A..29S}
}

@BOOK{osterbrock2006,
       author = {{Osterbrock}, Donald E. and {Ferland}, Gary J.},
        title = "{Astrophysics of gaseous nebulae and active galactic nuclei}",
         year = 2006,
       adsurl = {https://ui.adsabs.harvard.edu/abs/2006agna.book.....O}
}

@ARTICLE{kawabata2007,
       author = {{Kawabata}, K.~S. and {Ikeda}, Y. and {Akitaya}, H. and {Isogai}, M. and {Matsuda}, K. and {Matsumura}, M. and {Nagae}, O. and {Seki}, M.},
        title = "{Spectropolarimetry of R Coronae Borealis in 1998-2003: Discovery of Transient Polarization at Maximum Brightness}",
      journal = {\aj},
         year = 2007,
        month = nov,
       volume = {134},
       number = {5},
        pages = {1877-1889},
          doi = {10.1086/522629},
archivePrefix = {arXiv},
       eprint = {0708.2135},
 primaryClass = {astro-ph},
       adsurl = {https://ui.adsabs.harvard.edu/abs/2007AJ....134.1877K}
}

@ARTICLE{kumar2019,
       author = {{Kumar}, Brajesh and {Eswaraiah}, Chakali and {Singh}, Avinash and {Sahu}, D.~K. and {Anupama}, G.~C. and {Kawabata}, K.~S. and {Yamanaka}, Masayuki and {Otsubo}, Ikki and {Pandey}, S.~B. and {Nakaoka}, Tatsuya and {Kawabata}, Miho and {Aryan}, Amar and {Akitaya}, Hiroshi},
        title = "{On the observational behaviour of the highly polarized Type IIn supernova SN 2017hcc}",
      journal = {\mnras},
         year = 2019,
        month = sep,
       volume = {488},
       number = {3},
        pages = {3089-3099},
          doi = {10.1093/mnras/stz1914},
archivePrefix = {arXiv},
       eprint = {1907.03160},
 primaryClass = {astro-ph.HE},
       adsurl = {https://ui.adsabs.harvard.edu/abs/2019MNRAS.488.3089K}
}

@ARTICLE{moriya2013,
       author = {{Moriya}, Takashi J. and {Maeda}, Keiichi and {Taddia}, Francesco and {Sollerman}, Jesper and {Blinnikov}, Sergei I. and {Sorokina}, Elena I.},
        title = "{An analytic bolometric light curve model of interaction-powered supernovae and its application to Type IIn supernovae}",
      journal = {\mnras},
         year = 2013,
        month = oct,
       volume = {435},
       number = {2},
        pages = {1520-1535},
          doi = {10.1093/mnras/stt1392},
archivePrefix = {arXiv},
       eprint = {1307.2644},
 primaryClass = {astro-ph.HE},
       adsurl = {https://ui.adsabs.harvard.edu/abs/2013MNRAS.435.1520M}
}

@INCOLLECTION{smith2017,
       author = {{Smith}, Nathan},
        title = "{Interacting Supernovae: Types IIn and Ibn}",
    booktitle = {Handbook of Supernovae},
         year = 2017,
       editor = {{Alsabti}, Athem W. and {Murdin}, Paul},
        pages = {403},
          doi = {10.1007/978-3-319-21846-5_38},
       adsurl = {https://ui.adsabs.harvard.edu/abs/2017hsn..book..403S}
}

@ARTICLE{smith2024,
       author = {{Smith}, Nathan and {Andrews}, Jennifer E. and {Milne}, Peter and {Filippenko}, Alexei V. and {Brink}, Thomas G. and {Kelly}, Patrick L. and {Yuk}, Heechan and {Jencson}, Jacob E.},
        title = "{SN 2015da: late-time observations of a persistent superluminous Type IIn supernova with post-shock dust formation}",
      journal = {\mnras},
         year = 2024,
        month = may,
       volume = {530},
       number = {1},
        pages = {405-423},
          doi = {10.1093/mnras/stae726},
archivePrefix = {arXiv},
       eprint = {2312.00253},
 primaryClass = {astro-ph.HE},
       adsurl = {https://ui.adsabs.harvard.edu/abs/2024MNRAS.530..405S}
}

@ARTICLE{sarangi2018,
       author = {{Sarangi}, Arkaprabha and {Dwek}, Eli and {Arendt}, Richard G.},
        title = "{Delayed Shock-induced Dust Formation in the Dense Circumstellar Shell Surrounding the Type IIn Supernova SN 2010jl}",
      journal = {\apj},
         year = 2018,
        month = may,
       volume = {859},
       number = {1},
          eid = {66},
        pages = {66},
          doi = {10.3847/1538-4357/aabfc3},
archivePrefix = {arXiv},
       eprint = {1804.06878},
 primaryClass = {astro-ph.SR},
       adsurl = {https://ui.adsabs.harvard.edu/abs/2018ApJ...859...66S}
}

@ARTICLE{dwek2019,
       author = {{Dwek}, Eli and {Sarangi}, Arkaprabha and {Arendt}, Richard G.},
        title = "{The Evolution of Dust Opacity in Core Collapse Supernovae and the Rapid Formation of Dust in Their Ejecta}",
      journal = {\apjl},
         year = 2019,
        month = feb,
       volume = {871},
       number = {2},
          eid = {L33},
        pages = {L33},
          doi = {10.3847/2041-8213/aaf9a8},
archivePrefix = {arXiv},
       eprint = {1812.08234},
 primaryClass = {astro-ph.SR},
       adsurl = {https://ui.adsabs.harvard.edu/abs/2019ApJ...871L..33D}
}

@ARTICLE{fox2011,
       author = {{Fox}, Ori D. and {Chevalier}, Roger A. and {Skrutskie}, Michael F. and {Soderberg}, Alicia M. and {Filippenko}, Alexei V. and {Ganeshalingam}, Mohan and {Silverman}, Jeffrey M. and {Smith}, Nathan and {Steele}, Thea N.},
        title = "{A Spitzer Survey for Dust in Type IIn Supernovae}",
      journal = {\apj},
         year = 2011,
        month = nov,
       volume = {741},
       number = {1},
          eid = {7},
        pages = {7},
          doi = {10.1088/0004-637X/741/1/7},
archivePrefix = {arXiv},
       eprint = {1104.5012},
 primaryClass = {astro-ph.SR},
       adsurl = {https://ui.adsabs.harvard.edu/abs/2011ApJ...741....7F}
}

@ARTICLE{gall2014,
       author = {{Gall}, Christa and {Hjorth}, Jens and {Watson}, Darach and {Dwek}, Eli and {Maund}, Justyn R. and {Fox}, Ori and {Leloudas}, Giorgos and {Malesani}, Daniele and {Day-Jones}, Avril C.},
        title = "{Rapid formation of large dust grains in the luminous supernova 2010jl}",
      journal = {\nat},
         year = 2014,
        month = jul,
       volume = {511},
       number = {7509},
        pages = {326-329},
          doi = {10.1038/nature13558},
archivePrefix = {arXiv},
       eprint = {1407.4447},
 primaryClass = {astro-ph.SR},
       adsurl = {https://ui.adsabs.harvard.edu/abs/2014Natur.511..326G}
}

@ARTICLE{kozasa1989,
       author = {{Kozasa}, Takashi and {Hasegawa}, Hiroichi and {Nomoto}, Ken'ichi},
        title = "{Formation of Dust Grains in the Ejecta of SN 1987A}",
      journal = {\apj},
         year = 1989,
        month = sep,
       volume = {344},
        pages = {325},
          doi = {10.1086/167801},
       adsurl = {https://ui.adsabs.harvard.edu/abs/1989ApJ...344..325K}
}

@ARTICLE{wooden1993,
       author = {{Wooden}, Diane H. and {Rank}, David M. and {Bregman}, Jesse D. and {Witteborn}, Fred C. and {Tielens}, A.~G.~G.~M. and {Cohen}, Martin and {Pinto}, Philip A. and {Axelrod}, Timothy S.},
        title = "{Airborne Spectrophotometry of SN 1987A from 1.7 to 12.6 Microns: Time History of the Dust Continuum and Line Emission}",
      journal = {\apjs},
         year = 1993,
        month = oct,
       volume = {88},
        pages = {477},
          doi = {10.1086/191830},
       adsurl = {https://ui.adsabs.harvard.edu/abs/1993ApJS...88..477W}
}

@ARTICLE{sarangi2022,
       author = {{Sarangi}, Arkaprabha and {Slavin}, Jonathan D.},
        title = "{Dust Production in a Thin Dense Shell in Supernovae with Early Circumstellar Interactions}",
      journal = {\apj},
         year = 2022,
        month = jul,
       volume = {933},
       number = {1},
          eid = {89},
        pages = {89},
          doi = {10.3847/1538-4357/ac713d},
archivePrefix = {arXiv},
       eprint = {2205.08352},
 primaryClass = {astro-ph.SR},
       adsurl = {https://ui.adsabs.harvard.edu/abs/2022ApJ...933...89S}
}

@ARTICLE{smith2012,
       author = {{Smith}, Nathan and {Silverman}, Jeffrey M. and {Filippenko}, Alexei V. and {Cooper}, Michael C. and {Matheson}, Thomas and {Bian}, Fuyan and {Weiner}, Benjamin J. and {Comerford}, Julia M.},
        title = "{Systematic Blueshift of Line Profiles in the Type IIn Supernova 2010jl: Evidence for Post-shock Dust Formation?}",
      journal = {\aj},
         year = 2012,
        month = jan,
       volume = {143},
       number = {1},
          eid = {17},
        pages = {17},
          doi = {10.1088/0004-6256/143/1/17},
archivePrefix = {arXiv},
       eprint = {1108.2869},
 primaryClass = {astro-ph.HE},
       adsurl = {https://ui.adsabs.harvard.edu/abs/2012AJ....143...17S}
}

@ARTICLE{chugai2018,
       author = {{Chugai}, Nikolai N.},
        title = "{Type IIn SN 2010jl: probing dusty line-emitting shell}",
      journal = {\mnras},
         year = 2018,
        month = dec,
       volume = {481},
       number = {3},
        pages = {3643-3650},
          doi = {10.1093/mnras/sty2386},
archivePrefix = {arXiv},
       eprint = {1809.02478},
 primaryClass = {astro-ph.HE},
       adsurl = {https://ui.adsabs.harvard.edu/abs/2018MNRAS.481.3643C}
}

@ARTICLE{tinyanont2019,
       author = {{Tinyanont}, Samaporn and {Kasliwal}, Mansi M. and {Krafton}, Kelsie and {Lau}, Ryan and {Rho}, Jeonghee and {Leonard}, Douglas C. and {De}, Kishalay and {Jencson}, Jacob and {Mawet}, Dimitri and {Millar-Blanchaer}, Maxwell and {Nilsson}, Ricky and {Yan}, Lin and {Gehrz}, Robert D. and {Helou}, George and {Van Dyk}, Schuyler D. and {Serabyn}, Eugene and {Fox}, Ori D. and {Clayton}, Geoffrey},
        title = "{Supernova 2017eaw: Molecule and Dust Formation from Infrared Observations}",
      journal = {\apj},
         year = 2019,
        month = mar,
       volume = {873},
       number = {2},
          eid = {127},
        pages = {127},
          doi = {10.3847/1538-4357/ab0897},
archivePrefix = {arXiv},
       eprint = {1901.01940},
 primaryClass = {astro-ph.HE},
       adsurl = {https://ui.adsabs.harvard.edu/abs/2019ApJ...873..127T}
}

@ARTICLE{filippenko1997,
       author = {{Filippenko}, Alexei V.},
        title = "{Optical Spectra of Supernovae}",
      journal = {\araa},
         year = 1997,
        month = jan,
       volume = {35},
        pages = {309-355},
          doi = {10.1146/annurev.astro.35.1.309},
       adsurl = {https://ui.adsabs.harvard.edu/abs/1997ARA&A..35..309F}
}

@ARTICLE{schlegel1990,
       author = {{Schlegel}, Eric M.},
        title = "{A new subclass of type II supernovae ?}",
      journal = {\mnras},
         year = 1990,
        month = may,
       volume = {244},
        pages = {269-271},
       adsurl = {https://ui.adsabs.harvard.edu/abs/1990MNRAS.244..269S}
}

@ARTICLE{kiewe2012,
       author = {{Kiewe}, Michael and {Gal-Yam}, Avishay and {Arcavi}, Iair and {Leonard}, Douglas C. and {Emilio Enriquez}, J. and {Cenko}, S. Bradley and {Fox}, Derek B. and {Moon}, Dae-Sik and {Sand}, David J. and {Soderberg}, Alicia M. and {CCCP}, The},
        title = "{Caltech Core-Collapse Project (CCCP) Observations of Type IIn Supernovae: Typical Properties and Implications for Their Progenitor Stars}",
      journal = {\apj},
         year = 2012,
        month = jan,
       volume = {744},
       number = {1},
          eid = {10},
        pages = {10},
          doi = {10.1088/0004-637X/744/1/10},
archivePrefix = {arXiv},
       eprint = {1010.2689},
 primaryClass = {astro-ph.CO},
       adsurl = {https://ui.adsabs.harvard.edu/abs/2012ApJ...744...10K}
}

@ARTICLE{patat2011,
       author = {{Patat}, F. and {Taubenberger}, S. and {Benetti}, S. and {Pastorello}, A. and {Harutyunyan}, A.},
        title = "{Asymmetries in the type IIn SN 2010jl}",
      journal = {\aap},
         year = 2011,
        month = mar,
       volume = {527},
          eid = {L6},
        pages = {L6},
          doi = {10.1051/0004-6361/201016217},
archivePrefix = {arXiv},
       eprint = {1011.5926},
 primaryClass = {astro-ph.CO},
       adsurl = {https://ui.adsabs.harvard.edu/abs/2011A&A...527L...6P}
}

@ARTICLE{fraser2020,
       author = {{Fraser}, Morgan},
        title = "{Supernovae and transients with circumstellar interaction}",
      journal = {Royal Society Open Science},
         year = 2020,
        month = jul,
       volume = {7},
       number = {7},
          eid = {200467},
        pages = {200467},
          doi = {10.1098/rsos.200467},
       adsurl = {https://ui.adsabs.harvard.edu/abs/2020RSOS....700467F}
}

@ARTICLE{smith2009,
       author = {{Smith}, Nathan and {Silverman}, Jeffrey M. and {Chornock}, Ryan and {Filippenko}, Alexei V. and {Wang}, Xiaofeng and {Li}, Weidong and {Ganeshalingam}, Mohan and {Foley}, Ryan J. and {Rex}, Jacob and {Steele}, Thea N.},
        title = "{Coronal Lines and Dust Formation in SN 2005ip: Not the Brightest, but the Hottest Type IIn Supernova}",
      journal = {\apj},
         year = 2009,
        month = apr,
       volume = {695},
       number = {2},
        pages = {1334-1350},
          doi = {10.1088/0004-637X/695/2/1334},
archivePrefix = {arXiv},
       eprint = {0809.5079},
 primaryClass = {astro-ph},
       adsurl = {https://ui.adsabs.harvard.edu/abs/2009ApJ...695.1334S}
}

@ARTICLE{shahbandeh2025,
       author = {{Shahbandeh}, Melissa and {Fox}, Ori D. and {Temim}, Tea and {Dwek}, Eli and {Sarangi}, Arkaprabha and {Smith}, Nathan and {Dessart}, Luc and {Nickson}, Bryony and {Engesser}, Michael and {Filippenko}, Alexei V. and {Brink}, Thomas G. and {Zheng}, WeiKang and {Szalai}, Tam{\'a}s and {Johansson}, Joel and {Rest}, Armin and {Van Dyk}, Schuyler D. and {Andrews}, Jennifer and {Ashall}, Chris and {Clayton}, Geoffrey C. and {De Looze}, Ilse and {DerKacy}, James M. and {Dulude}, Michael and {Foley}, Ryan J. and {Gezari}, Suvi and {Gomez}, Sebastian and {Gonzaga}, Shireen and {Indukuri}, Siva and {Jencson}, Jacob and {Kasliwal}, Mansi and {Lane}, Zachary G. and {Lau}, Ryan and {Law}, David and {Marston}, Anthony and {Milisavljevic}, Dan and {O'Steen}, Richard and {Pierel}, Justin and {Siebert}, Matthew and {Skrutskie}, Michael and {Strolger}, Lou and {Tinyanont}, Samaporn and {Wang}, Qinan and {Williams}, Brian and {Xiao}, Lin and {Yang}, Yi and {Zs{\'\i}ros}, Szanna},
        title = "{JWST/MIRI Observations of Newly Formed Dust in the Cold, Dense Shell of the Type IIn SN 2005ip}",
      journal = {\apj},
         year = 2025,
        month = jun,
       volume = {985},
       number = {2},
          eid = {262},
        pages = {262},
          doi = {10.3847/1538-4357/adce77},
archivePrefix = {arXiv},
       eprint = {2410.09142},
 primaryClass = {astro-ph.HE},
       adsurl = {https://ui.adsabs.harvard.edu/abs/2025ApJ...985..262S}
}

@ARTICLE{suzuki2019,
       author = {{Suzuki}, Akihiro and {Moriya}, Takashi J. and {Takiwaki}, Tomoya},
        title = "{Supernova Ejecta Interacting with a Circumstellar Disk. I. Two-dimensional Radiation-hydrodynamic Simulations}",
      journal = {\apj},
         year = 2019,
        month = dec,
       volume = {887},
       number = {2},
          eid = {249},
        pages = {249},
          doi = {10.3847/1538-4357/ab5a83},
archivePrefix = {arXiv},
       eprint = {1911.09261},
 primaryClass = {astro-ph.HE},
       adsurl = {https://ui.adsabs.harvard.edu/abs/2019ApJ...887..249S}
}

@ARTICLE{ivezi2019,
       author = {{Ivezi{\'c}}, {\v{Z}}eljko and {Kahn}, Steven M. and {Tyson}, J. Anthony and {Abel}, Bob and {Acosta}, Emily and {Allsman}, Robyn and {Alonso}, David and {AlSayyad}, Yusra and {Anderson}, Scott F. and {Andrew}, John and {Angel}, James Roger P. and {Angeli}, George Z. and {Ansari}, Reza and {Antilogus}, Pierre and {Araujo}, Constanza and {Armstrong}, Robert and {Arndt}, Kirk T. and {Astier}, Pierre and {Aubourg}, {\'E}ric and {Auza}, Nicole and {Axelrod}, Tim S. and {Bard}, Deborah J. and {Barr}, Jeff D. and {Barrau}, Aurelian and {Bartlett}, James G. and {Bauer}, Amanda E. and {Bauman}, Brian J. and {Baumont}, Sylvain and {Bechtol}, Ellen and {Bechtol}, Keith and {Becker}, Andrew C. and {Becla}, Jacek and {Beldica}, Cristina and {Bellavia}, Steve and {Bianco}, Federica B. and {Biswas}, Rahul and {Blanc}, Guillaume and {Blazek}, Jonathan and {Blandford}, Roger D. and {Bloom}, Josh S. and {Bogart}, Joanne and {Bond}, Tim W. and {Booth}, Michael T. and {Borgland}, Anders W. and {Borne}, Kirk and {Bosch}, James F. and {Boutigny}, Dominique and {Brackett}, Craig A. and {Bradshaw}, Andrew and {Brandt}, William Nielsen and {Brown}, Michael E. and {Bullock}, James S. and {Burchat}, Patricia and {Burke}, David L. and {Cagnoli}, Gianpietro and {Calabrese}, Daniel and {Callahan}, Shawn and {Callen}, Alice L. and {Carlin}, Jeffrey L. and {Carlson}, Erin L. and {Chandrasekharan}, Srinivasan and {Charles-Emerson}, Glenaver and {Chesley}, Steve and {Cheu}, Elliott C. and {Chiang}, Hsin-Fang and {Chiang}, James and {Chirino}, Carol and {Chow}, Derek and {Ciardi}, David R. and {Claver}, Charles F. and {Cohen-Tanugi}, Johann and {Cockrum}, Joseph J. and {Coles}, Rebecca and {Connolly}, Andrew J. and {Cook}, Kem H. and {Cooray}, Asantha and {Covey}, Kevin R. and {Cribbs}, Chris and {Cui}, Wei and {Cutri}, Roc and {Daly}, Philip N. and {Daniel}, Scott F. and {Daruich}, Felipe and {Daubard}, Guillaume and {Daues}, Greg and {Dawson}, William and {Delgado}, Francisco and {Dellapenna}, Alfred and {de Peyster}, Robert and {de Val-Borro}, Miguel and {Digel}, Seth W. and {Doherty}, Peter and {Dubois}, Richard and {Dubois-Felsmann}, Gregory P. and {Durech}, Josef and {Economou}, Frossie and {Eifler}, Tim and {Eracleous}, Michael and {Emmons}, Benjamin L. and {Fausti Neto}, Angelo and {Ferguson}, Henry and {Figueroa}, Enrique and {Fisher-Levine}, Merlin and {Focke}, Warren and {Foss}, Michael D. and {Frank}, James and {Freemon}, Michael D. and {Gangler}, Emmanuel and {Gawiser}, Eric and {Geary}, John C. and {Gee}, Perry and {Geha}, Marla and {Gessner}, Charles J.~B. and {Gibson}, Robert R. and {Gilmore}, D. Kirk and {Glanzman}, Thomas and {Glick}, William and {Goldina}, Tatiana and {Goldstein}, Daniel A. and {Goodenow}, Iain and {Graham}, Melissa L. and {Gressler}, William J. and {Gris}, Philippe and {Guy}, Leanne P. and {Guyonnet}, Augustin and {Haller}, Gunther and {Harris}, Ron and {Hascall}, Patrick A. and {Haupt}, Justine and {Hernandez}, Fabio and {Herrmann}, Sven and {Hileman}, Edward and {Hoblitt}, Joshua and {Hodgson}, John A. and {Hogan}, Craig and {Howard}, James D. and {Huang}, Dajun and {Huffer}, Michael E. and {Ingraham}, Patrick and {Innes}, Walter R. and {Jacoby}, Suzanne H. and {Jain}, Bhuvnesh and {Jammes}, Fabrice and {Jee}, M. James and {Jenness}, Tim and {Jernigan}, Garrett and {Jevremovi{\'c}}, Darko and {Johns}, Kenneth and {Johnson}, Anthony S. and {Johnson}, Margaret W.~G. and {Jones}, R. Lynne and {Juramy-Gilles}, Claire and {Juri{\'c}}, Mario and {Kalirai}, Jason S. and {Kallivayalil}, Nitya J. and {Kalmbach}, Bryce and {Kantor}, Jeffrey P. and {Karst}, Pierre and {Kasliwal}, Mansi M. and {Kelly}, Heather and {Kessler}, Richard and {Kinnison}, Veronica and {Kirkby}, David and {Knox}, Lloyd and {Kotov}, Ivan V. and {Krabbendam}, Victor L. and {Krughoff}, K. Simon and {Kub{\'a}nek}, Petr and {Kuczewski}, John and {Kulkarni}, Shri and {Ku}, John and {Kurita}, Nadine R. and {Lage}, Craig S. and {Lambert}, Ron and {Lange}, Travis and {Langton}, J. Brian and {Le Guillou}, Laurent and {Levine}, Deborah and {Liang}, Ming and {Lim}, Kian-Tat and {Lintott}, Chris J. and {Long}, Kevin E. and {Lopez}, Margaux and {Lotz}, Paul J. and {Lupton}, Robert H. and {Lust}, Nate B. and {MacArthur}, Lauren A. and {Mahabal}, Ashish and {Mandelbaum}, Rachel and {Markiewicz}, Thomas W. and {Marsh}, Darren S. and {Marshall}, Philip J. and {Marshall}, Stuart and {May}, Morgan and {McKercher}, Robert and {McQueen}, Michelle and {Meyers}, Joshua and {Migliore}, Myriam and {Miller}, Michelle and {Mills}, David J.},
        title = "{LSST: From Science Drivers to Reference Design and Anticipated Data Products}",
      journal = {\apj},
         year = 2019,
        month = mar,
       volume = {873},
       number = {2},
          eid = {111},
        pages = {111},
          doi = {10.3847/1538-4357/ab042c},
archivePrefix = {arXiv},
       eprint = {0805.2366},
 primaryClass = {astro-ph},
       adsurl = {https://ui.adsabs.harvard.edu/abs/2019ApJ...873..111I}
}

@ARTICLE{jacobson2025,
       author = {{Jacobson-Gal{\'a}n}, W.~V. and {Dessart}, L. and {Kilpatrick}, C.~D. and {Patel}, P.~J. and {Auchettl}, K. and {Tinyanont}, S. and {Margutti}, R. and {Dwarkadas}, V.~V. and {Bostroem}, K.~A. and {Chornock}, R. and {Foley}, R.~J. and {Abunemeh}, H. and {Ahumada}, T. and {Arunachalam}, P. and {Bustamante-Rosell}, M.~J. and {Coulter}, D.~A. and {Gall}, C. and {Gao}, H. and {Guo}, X. and {Jones}, D.~O. and {Hjorth}, J. and {Kaewmookda}, M. and {Kasliwal}, M.~M. and {Kaur}, R. and {Larison}, C. and {LeBaron}, N. and {Miao}, H.-Y. and {Narayan}, G. and {Pan}, Y.-C. and {Park}, S.~H. and {Patra}, K.~C. and {Qin}, Y. and {Ransome}, C.~L. and {Rest}, A. and {Rho}, J. and {Rose}, S. and {Sears}, H. and {Swift}, J.~J. and {Taggart}, K. and {Villar}, V.~A. and {Wang}, Q. and {Zenati}, Y. and {Zhou}, H.},
        title = "{A Panchromatic View of Late-time Shock Power in the Type II Supernova 2023ixf}",
      journal = {\apjl},
         year = 2025,
        month = nov,
       volume = {994},
       number = {1},
          eid = {L14},
        pages = {L14},
          doi = {10.3847/2041-8213/ae157a},
archivePrefix = {arXiv},
       eprint = {2508.11747},
 primaryClass = {astro-ph.HE},
       adsurl = {https://ui.adsabs.harvard.edu/abs/2025ApJ...994L..14J}
}

@ARTICLE{dessart2025,
       author = {{Dessart}, Luc and {John Hillier}, D. and {Sarangi}, Arkaprabha},
        title = "{Radiative-transfer models for dusty Type II supernovae}",
      journal = {\aap},
         year = 2025,
        month = jun,
       volume = {698},
          eid = {A293},
        pages = {A293},
          doi = {10.1051/0004-6361/202555161},
archivePrefix = {arXiv},
       eprint = {2504.10928},
 primaryClass = {astro-ph.SR},
       adsurl = {https://ui.adsabs.harvard.edu/abs/2025A&A...698A.293D}
}

@ARTICLE{mattila2008,
       author = {{Mattila}, S. and {Meikle}, W.~P.~S. and {Lundqvist}, P. and {Pastorello}, A. and {Kotak}, R. and {Eldridge}, J. and {Smartt}, S. and {Adamson}, A. and {Gerardy}, C.~L. and {Rizzi}, L. and {Stephens}, A.~W. and {van Dyk}, S.~D.},
        title = "{Massive stars exploding in a He-rich circumstellar medium - III. SN 2006jc: infrared echoes from new and old dust in the progenitor CSM}",
      journal = {\mnras},
         year = 2008,
        month = sep,
       volume = {389},
       number = {1},
        pages = {141-155},
          doi = {10.1111/j.1365-2966.2008.13516.x},
archivePrefix = {arXiv},
       eprint = {0803.2145},
 primaryClass = {astro-ph},
       adsurl = {https://ui.adsabs.harvard.edu/abs/2008MNRAS.389..141M}
}

@ARTICLE{tonry2024TNS,
       author = {{Tonry}, J. and {Denneau}, L. and {Weiland}, H. and {Lawrence}, A. and {Siverd}, R. and {Erasmus}, N. and {Koorts}, W. and {Jordan}, A. and {Suc}, V. and {Smartt}, S.~J. and {Smith}, K.~W. and {Young}, D.~R. and {Nicholl}, M. and {Fulton}, M. and {McCollum}, M. and {Moore}, T. and {Weston}, J. and {Sheng}, X. and {Ramsden}, P. and {Aamer}, A. and {Shingles}, L. and {Srivastav}, S. and {Ramaiya}, S. and {Gillanders}, J. and {Rhodes}, L. and {Andersson}, A. and {Stevance}, H. and {Rest}, A. and {Chen}, T.~W. and {Stubbs}, C. and {Sommer}, J.},
        title = "{ATLAS Transient Discovery Report for 2024-01-03}",
      journal = {Transient Name Server Discovery Report},
         year = 2024,
        month = jan,
       volume = {2024-17},
        pages = {1},
       adsurl = {https://ui.adsabs.harvard.edu/abs/2024TNSTR..17....1T}
}

@ARTICLE{moriya2020,
       author = {{Moriya}, T.~J. and {Stritzinger}, M.~D. and {Taddia}, F. and {Morrell}, N. and {Suntzeff}, N.~B. and {Contreras}, C. and {Gall}, C. and {Hjorth}, J. and {Ashall}, C. and {Burns}, C.~R. and {Busta}, L. and {Campillay}, A. and {Castell{\'o}n}, S. and {Corco}, C. and {Davis}, S. and {Galbany}, L. and {Gonz{\'a}lez}, C. and {Holmbo}, S. and {Hsiao}, E.~Y. and {Maund}, J.~R. and {Phillips}, M.~M.},
        title = "{The Carnegie Supernova Project II. Observations of SN 2014ab possibly revealing a 2010jl-like SN IIn with pre-existing dust}",
      journal = {\aap},
         year = 2020,
        month = sep,
       volume = {641},
          eid = {A148},
        pages = {A148},
          doi = {10.1051/0004-6361/202038118},
archivePrefix = {arXiv},
       eprint = {2006.10198},
 primaryClass = {astro-ph.HE},
       adsurl = {https://ui.adsabs.harvard.edu/abs/2020A&A...641A.148M}
}

@ARTICLE{rouleau1991,
       author = {{Rouleau}, Francois and {Martin}, P.~G.},
        title = "{Shape and Clustering Effects on the Optical Properties of Amorphous Carbon}",
      journal = {\apj},
         year = 1991,
        month = aug,
       volume = {377},
        pages = {526},
          doi = {10.1086/170382},
       adsurl = {https://ui.adsabs.harvard.edu/abs/1991ApJ...377..526R}
}

@ARTICLE{draine1985,
       author = {{Draine}, B.~T.},
        title = "{Tabulated Optical Properties of Graphite and Silicate Grains}",
      journal = {\apjs},
         year = 1985,
        month = mar,
       volume = {57},
        pages = {587},
          doi = {10.1086/191016},
       adsurl = {https://ui.adsabs.harvard.edu/abs/1985ApJS...57..587D}
}

@ARTICLE{szalai2021,
       author = {{Szalai}, Tam{\'a}s and {Fox}, Ori D. and {Arendt}, Richard G. and {Dwek}, Eli and {Andrews}, Jennifer E. and {Clayton}, Geoffrey C. and {Filippenko}, Alexei V. and {Johansson}, Joel and {Kelly}, Patrick L. and {Krafton}, Kelsie and {Marston}, A.~P. and {Mauerhan}, Jon C. and {Van Dyk}, Schuyler D.},
        title = "{Spitzer's Last Look at Extragalactic Explosions: Long-term Evolution of Interacting Supernovae}",
      journal = {\apj},
         year = 2021,
        month = sep,
       volume = {919},
       number = {1},
          eid = {17},
        pages = {17},
          doi = {10.3847/1538-4357/ac0e2b},
archivePrefix = {arXiv},
       eprint = {2106.12427},
 primaryClass = {astro-ph.HE},
       adsurl = {https://ui.adsabs.harvard.edu/abs/2021ApJ...919...17S}
}

@ARTICLE{bevan2019,
       author = {{Bevan}, A. and {Wesson}, R. and {Barlow}, M.~J. and {De Looze}, I. and {Andrews}, J.~E. and {Clayton}, G.~C. and {Krafton}, K. and {Matsuura}, M. and {Milisavljevic}, D.},
        title = "{A decade of ejecta dust formation in the Type IIn SN 2005ip}",
      journal = {\mnras},
         year = 2019,
        month = jun,
       volume = {485},
       number = {4},
        pages = {5192-5206},
          doi = {10.1093/mnras/stz679},
archivePrefix = {arXiv},
       eprint = {1809.09055},
 primaryClass = {astro-ph.SR},
       adsurl = {https://ui.adsabs.harvard.edu/abs/2019MNRAS.485.5192B}
}

@ARTICLE{dwek1985,
       author = {{Dwek}, E.},
        title = "{The infrared echo of type II supernovae with circumstellar dust shells. II. A probe into the presupernova evolution of the progenitor star.}",
      journal = {\apj},
         year = 1985,
        month = oct,
       volume = {297},
        pages = {719-723},
          doi = {10.1086/163571},
       adsurl = {https://ui.adsabs.harvard.edu/abs/1985ApJ...297..719D}
}
\bibliographystyle{aasjournalv7}

\end{document}